\documentclass[fleqn,usenatbib]{mnras}

\usepackage[T1]{fontenc}

\DeclareRobustCommand{\VAN}[3]{#2}
\let\VANthebibliography\thebibliography
\def\thebibliography{\DeclareRobustCommand{\VAN}[3]{##3}\VANthebibliography}

\usepackage{graphicx}	
\usepackage{amsmath}	
\usepackage{amssymb}	
\usepackage{placeins}
\usepackage{color}
\usepackage{bm}
\usepackage{xspace}

\newcommand{\lae}{{\rm LAE}}
\newcommand{\oii}{[\ion{O}{II}]\xspace}
\newcommand{\oiii}{[\ion{O}{III}]\xspace}
\newcommand{\true}{\mathrm{true}}
\newcommand{\est}{\mathrm{est}}
\newcommand{\LC}{\mathrm{LC}}
\newcommand{\proj}{\mathrm{proj}}
\newcommand{\obs}{\mathrm{obs}}

\title[Redshift dependent contamination]{Correcting correlation functions for redshift-dependent interloper contamination}

\author[Daniel J. Farrow et al.]{Daniel J. Farrow,$^{1, 2}$ Ariel G. S\'anchez,$^{1, 2}$, Robin Ciardullo$^{3,4}$, Erin Mentuch Cooper$^{5}$, Dustin Davis$^{5}$,\newauthor Maximilian Fabricius,$^{1,2}$, Eric Gawiser,$^{6}$   
Henry S. Grasshorn Gebhardt,$^{7,8}$ Karl Gebhardt,$^{5}$\newauthor 
Gary J. Hill,$^{5,9}$ Donghui Jeong,$^{3, 4}$ Eiichiro Komatsu,$^{10,11}$ Martin Landriau,$^{12}$ Chenxu Liu,$^{5}$\newauthor Shun Saito,$^{13,11}$
Jan Snigula,$^{1,2}$ Isak G. B. Wold$^{14}$
\\
$^{1}$Max-Planck-Institut f{\"u}r extraterrestrische Physik, Giessenbachstrasse 1, 85748 Garching, Germany\\
$^{2}$Universit{\"a}ts-Sternwarte, Fakult{\"a}t f{\"u}r Physik, Ludwig-Maximilians-Universit{\"a}t M{\"u}nchen, Scheinerstr. 1, 81679  M{\"u}nchen, Germany \\
$^{3}$Department of Astronomy \& Astrophysics, The Pennsylvania State University, University Park, PA 169802, USA\\
$^{4}$Institute for Gravitation \& the Cosmos, The Pennsylvania State University, University Park, PA 169802, USA\\
$^{5}$Department of Astronomy, University of Texas at Austin, 2515 Speedway, Stop C1400, Austin, Texas 78712, USA\\
$^{6}$Rutgers, The State University of New Jersey, Piscataway, NJ 08854, USA\\
$^{7}$Jet Propulsion Laboratory, California Institute of Technology, Pasadena, CA 91109, USA \\
$^{8}$California Institute of Technology, Pasadena, CA 91125, USA \\
$^{9}$McDonald Observatory, University of Texas at Austin, 2515 Speedway, Stop C1402, Austin, TX 78712, USA\\
$^{10}$ Max-Planck-Institut f\"{u}r Astrophysik, Karl-Schwarzschild Str. 1, 85741 Garching, Germany\\
$^{11}$Kavli Institute for the Physics and Mathematics of the Universe (Kavli IPMU, WPI), University of Tokyo, Chiba 277-8582, Japan\\
$^{12}$Lawrence Berkeley National Laboratory, 1 Cyclotron Road, Berkeley, CA 94720, USA\\
$^{13}$Institute for Multi-messenger Astrophysics and Cosmology, Department of Physics, Missouri University of Science and Technology,\\1315 N Pine St, Rolla, MO 65409, USA\\
$^{14}$Astrophysics Science Division, NASA Goddard Space Flight Center, 8800 Greenbelt Road, Greenbelt, Maryland, 20771, USA\\
}

\date{Accepted XXX. Received YYY; in original form ZZZ}

\pubyear{2021}

\begin{document}
\label{firstpage}
\pagerange{\pageref{firstpage}--\pageref{lastpage}}
\maketitle

\begin{abstract}
The construction of catalogues of a particular type of galaxy can be complicated by interlopers contaminating
the sample. In spectroscopic galaxy surveys this can be due to the misclassification of an 
emission line; for example in the Hobby-Eberly Telescope Dark Energy Experiment (HETDEX) low redshift \oii emitters may make
up a few percent of the observed Ly~$\alpha$ emitter (LAE) sample. The presence of contaminants affects the measured
correlation functions and power spectra. Previous attempts to deal with this using the cross-correlation function have assumed
sources at a fixed redshift, or not modelled evolution within the adopted redshift bins. However, in spectroscopic surveys like HETDEX, 
where the contamination fraction is likely to be redshift dependent, the observed clustering of misclassified sources will appear to evolve strongly due to projection effects, even if their true clustering does not. We present a practical method for accounting for the presence of contaminants with redshift-dependent contamination 
fractions and projected clustering. We show using mock catalogues that our method, unlike existing approaches,
yields unbiased clustering measurements from the upcoming HETDEX survey in scenarios with redshift-dependent contamination fractions within the redshift bins used.
We show our method returns auto-correlation functions with systematic biases much smaller than the statistical noise for samples with at least as high as 7 per cent contamination. We also present and test a method for fitting for the redshift-dependent interloper fraction using the LAE-\oii galaxy cross-correlation function, which gives less biased results than assuming a single interloper fraction for the whole sample.
\end{abstract}




\section{Introduction}
The measurement of a redshift from a galaxy
spectrum is one of the most fundamental parts
of a spectroscopic survey. This is usually achieved by relying on features in the spectra such as emission and absorption lines and
the shape of the continuum.  However, when only
one emission line is detected it becomes impossible to unambiguously identify the rest frame emission line and return an accurate classification and redshift. This results in catalogues of galaxies which contain interlopers, i.e., misclassified sources at the wrong redshift. Interloper contamination is expected to be important in several major upcoming galaxy surveys \cite[for examples see e.g.][]{pullen2016}. The focus of this paper is the ongoing Hobby-Eberly Telescope Dark Energy Experiment (HETDEX;  \citealt{hill2008}, Hill et al in prep, Gebhardt et al in prep), where, due to the spectrographs not resolving the \oii\ doublet, low redshift \oii emitters with rest-frame wavelength 3727\,\AA\ can be mistaken for high redshift Ly~$\alpha$ emitters (LAEs) with rest-frame wavelength 1216\,\AA. 

The impact of interlopers on the correlation function and power spectrum of a galaxy sample has been studied in the literature \citep[e.g.,][]{pullen2016, leung, grasshorn19, addison2019, massara2020}. It has been seen that the presence of interlopers in a sample changes the galaxies' correlation function and power spectrum. It is also understood that if the interlopers are unclustered then the main effect just decreases the
overall clustering amplitude by adding in uncorrelated sources \citep[see Appendix B.4 of][]{grasshorn19}.  However, if the interlopers are clustered, then a signal from their correlation function is added into the sample. It has also been shown that these spurious clustering signals can cause biases in the inferred cosmological parameters \citep[e.g.][]{pullen2016, grasshorn19, addison2019}. 

In both \citet{grasshorn19} and \citet{addison2019} methods are presented that  include the effects of interlopers in the modelling of the galaxy power spectrum. These authors note that a cross correlation signal between two intrinsically uncorrelated samples of galaxies can be created entirely due to interloper contamination. They advocate using this observed cross correlation signal to put constraints on the contamination fraction, in order to yield better measurements of cosmological parameters. An alternative approach to forward modelling techniques is to decontaminate the measurements by applying a transformation that
changes the observed auto and cross 
correlation functions into the true underlying functions.  A matrix to carry out this transformation
and its inverse is given in \citet{awan20}. Their work deals with angular clustering measurements in redshift bins.

A related issue to interlopers in spectroscopic galaxy surveys is their impact in line intensity mapping experiments \citep[e.g.,][]{visbal2010, gong2014, gong2020, lidz2016, cheng2016, cheng2020}. These studies differ from emission line surveys in that they target the light from unresolved populations of galaxies.  However, it has also been noted that interlopers in intensity mapping experiments add an anisotropic signal to the power spectrum of the target population \citep[e.g.,][]{visbal2010,gong2014, lidz2016}. In \citet{gong2020} a method is presented that jointly fits the cosmology and properties of interloper lines in line intensity mapping experiments.

One scenario that has not been addressed by efforts to model the correlation function or power spectrum from spectroscopic emission line surveys is when the contamination fractions and the clustering of the contaminants show rapid evolution within the redshift bins used to define samples. Existing methods may work to an acceptable level with correlation functions that have a reasonable amount of evolution within the redshift bins considered, but in HETDEX the observed \oii clustering
signal will evolve rapidly with redshift, due to projection effects \citep[see e.g. Figure 2 of][]{grasshorn19}. The \oii contamination fraction will also be redshift dependent, due to the intrinsic redshift distribution of the emission lines and due to the wavelength dependence of the noise. Although \citet{cheng2020} recently published a method of generating a 3D lightcone of the interlopers in an intensity mapping survey, their method relies on the interlopers having multiple emission lines. That will not usually
be the case for HETDEX, as beyond $z \sim 0.13$, the bulk of the \oii galaxy population will only have a single detectable emission-line. \citet{cheng2020} also focuses on producing a 3D map of the interloper density, not unbiased correlation function measurements from the target population.

In this paper we present a method to account for the redshift dependence of the contamination fractions in emission-line surveys by combining the
decontamination methodology in the literature  with lightcone effects presented in \citet{yamamoto1999} and \citet{suto2000}. References to `lightcone effects' in this paper specifically refer to effects from the redshift dependent contamination and observed clustering. We test our method on simulations of the HETDEX survey, and demonstrate that our method
to deal with the lightcone effects is an improvement over assuming fixed contamination fractions and clustering across a whole redshift bin. We also show that
our new method is useful when using the cross-correlation function to gain  unbiased constraints on the contamination fractions. We focus on HETDEX here, but the work we present gives insights into all surveys with contamination rates that depend on redshift.

The outline of the paper is as follows: in section~\ref{sec:sims} we introduce the HETDEX survey and our simulations of it; this section also includes a method of assigning source classification probabilities. In section~\ref{sec:corrf} we present the methods used to measure and model the projected clustering. Then in section~\ref{sec:deconmed} we present the methodology of our decontamination.   We show the results of our model in section~\ref{sec:results}, and in section~\ref{sec:fit} we use our new methodology to fit for the redshift-dependent contamination. We give our conclusions in section~\ref{sec:conc}.

\section{Simulations of HETDEX}\label{sec:sims}

In this section we explain how we generate mock catalogues. We note that our work follows that of \citet{chiang2013}, who use an older version of the log-normal simulation code used here \citep{agrawal2017}, and an older HETDEX design, to produce simulations of the HETDEX survey. We improve on that paper, first by adding \oii galaxies and source classifications following \citet{leung}, and then by
adding in more realistic redshift dependent variations into the sensitivity and 
noise estimates.

We will begin by introducing HETDEX (section~\ref{sec:hsurvey}), then the following sections introduce the model of large-scale structure (section~\ref{sec:lssmodel}) and the approach we use to generate a density field with a given power spectrum (section~\ref{sec:lognorm}). We also explain how we assign galaxy properties (section~\ref{sec:lae_props} and \ref{sec:spectra}), model observational effects (sections~\ref{sec:alself}, \ref{sec:rsel} and \ref{sec:ext}) and assign the LAE probabilities (section \ref{sec:probs}) to generate samples of LAEs and [OII] emitters.

\subsection{The HETDEX survey}
\label{sec:hsurvey}
HETDEX
is a program on the Hobby-Eberly Telescope at the McDonald Observatory, Texas (\citealt{hill2008}; Hill et al in prep; Gebhardt et al in prep) to use LAEs to map out the large scale structure of the 
$1.9<z<3.5$ Universe.  The survey measures spectra from the sky using 
an array of up to 78 integral field units (IFUs; \citealt{hill2018}; Hill et al in prep), galaxies are not pre-selected but instead observations are taken blindly. Each IFU has a square footprint roughly 50\arcsec\ on a side, and neighbouring IFUs are separated by 100\arcsec. When observing, the gaps between the fibers are filled in by taking 3 dithered exposures. The dithering does not fill in the gaps between the IFUs, however, meaning areas of sky are sparsely sampled. It has been
shown that such a sampling can be treated as surveying the whole area with a lower number of tracers \citep{chiang2013}. We refer to the set of three dithers at one pointing as an `exposure set', and use the term `exposure set position' to refer to the right-ascension and declination of the pointing.

The survey sparsely samples two main fields: a roughly 390 deg$^2$ field in the northern hemisphere (the `Spring' field) and a $\sim  150$~deg$^2$ equatorial region (the `Fall' field). Defining the area that is sparsely sampled is difficult due to the jagged edges of the HETDEX footprint, which are caused by the approximately octagonal boundary of IFUs in the focal plane.  In the real survey additional effects we do not model here, such as bright stars in the Milky Way and large foreground galaxies  create holes in the survey, further complicating the issue. Thus,
the precise values for the survey areas depend on how survey edges are defined and the regions which are compromised by foreground sources. 

The survey goal is to measure the clustering (e.g., correlation function or power spectrum) of the LAEs and use it to probe cosmology. The modest
resolution of the spectrographs (mean resolving power $R= \lambda$/$\delta\lambda$ $\sim$ 800) means the \oii doublet cannot be resolved, resulting in some \oii emitters being classified as LAEs
\citep[see][]{leung}.
\begin{table*}
\centering
\begin{tabular}{|l l l l l|}
\hline
\multicolumn{2}{|l|}{Cosmology - flat $\Lambda$CDM \citep[][with a small modification, see caption]{planck2018}}\\
\hline
\hline
$H$ & 67.36 km s$^{-1}$ Mpc$^{-1}$ \\
$\Omega_{\mathrm{b}}h^2$ & 0.02237 \\
$\Omega_{\mathrm{c}}h^2$ &  0.12 \\
$\Omega_{\mathrm{k}}$ & 0 \\
$n_{\mathrm{s}}$ & 0.9649 \\
$\sigma_8$ & $0.8226$ \\
$\sigma_{12}$ & $0.8167$ \\
\hline
\multicolumn{3}{|l|}{LAE Luminosity and EW functions \citep{gronwall2014}}\\
\hline
\hline
Redshift   &    2.063   & 3.104 \\
\hline
$L^*(h=0.7)$[erg s$^{-1}]$ & $4.07 \times 10^{42}$ & $5.98 \times 10^{42}$ \\
$\phi^{*}(h=0.7)$[Mpc$^{-3}$] & $8.32 \times 10^{-4}$ & $1.05 \times 10^{-3}$ \\
$\alpha$ & -1.65 & -1.65\\
$w_0$ [\AA] & 50 & 100 \\
\hline
\multicolumn{2}{|l|}{\oii Luminosity and EW function \citep{ciardullo2013}} \\
\hline
\hline
Redshift  & 0.1 & 0.2625 & 0.3875 & 0.5050 \\
\hline
$L^*(h=0.7)$[erg s$^{-1}$] & $1.17 \times 10^{41}$ & $1.95 \times 10^{41}$ & $3.16 \times 10^{41}$  & $3.79 \times 10^{41}$  \\
  $\phi^{*}(h=0.7)$[Mpc$^{-3}$] & $5.01 \times 10^{-3}$ & $7.59 \times 10^{-3}$ & $8.51 \times 10^{-3}$ & $8.51 \times 10^{-3}$   \\
$\alpha$ & -1.2 & -1.2 & -1.2 & -1.2 \\
$w_0$ [\AA] & 8.00 & 11.5 & 16.6 & 21.5 \\
\hline
\multicolumn{2}{|l|}{Survey Properties (Sections~\ref{sec:alself} \& \ref{sec:rsel})}\\
\hline
\hline
Field & Spring & Fall \\
\hline
Total Area (with gaps) [deg$^2$] & 390  & 150 \\
Total Area (covered by fibers) [deg$^2$] & 55.6 &  27.2\\
Volume with LAEs [$h^{-3}$Gpc$^{3}$] & 2.42 & 0.93 \\
Number of IFUs & 78 & 78 \\
Number of LAEs & $6.4 \times 10^{5}$ & $2.9 \times 10^{5}$\\
Number of \oii galaxies & $4.2 \times 10^5$ & $2.0 \times 10^{5}$ \\
LAE Number density  [$h^{3}$Mpc$^{-3}$] & $2.7 \times 10^{-4}$  & $3.1 \times 10^{-4}$\\
\hline
\end{tabular}
\caption{A short summary of the important assumptions and input parameters for the mocks of an idealized HETDEX survey. The cosmological parameters are from \citet{planck2018}. The values for angular area of the survey are explained in section~\ref{sec:alself}, 
the prediction for the number of LAEs is explained in section~\ref{sec:rsel}\label{table:summary}. The volume
given is for the LAE redshift range and for the total area that is covered with gaps and sparse observations \citep[see][]{chiang2013}. The number density assumes the total number
of LAEs are spread uniformly over that volume. As we, unlike \citet{planck2018}, assume 
massless neutrinos, we do not use their quoted $\sigma_{8}$ value but instead compute it
using \citet{lewis2000}. We also include $\sigma_{12}$, the square root of the variance in 
12~Mpc spheres (i.e. not using $h$ units), as an alternative to the more standard 
$\sigma_{8}$ \citep[see the arguments in][]{sanchez2020}.} 
\end{table*}

\subsection{Model of Cosmology \& Large-Scale Structure}
\label{sec:lssmodel}
The simulations and our whole paper use the marginalised mean, flat $\Lambda$CDM cosmology from the \citet{planck2018}, but for simplicity we assume massless neutrinos (see table~\ref{table:summary} for the exact parameter values).
The model of the power spectrum and bias used to generate the  simulations is the same as that
used for the analysis of the Baryon Oscillation Spectroscopic Survey \citep[BOSS;][]{dawson2013} 
 by \citet{sanchez2017}, and a full 
description of the model can be found there. Briefly, a linear power spectrum is generated at the mean pair redshift\footnote[1]{the mean over all pairs of $(z_1 + z_2)/2$, where $z_1$ and $z_2$ are the redshifts of each galaxy in the pair.} of the \oii  ($z=0.3$) 
and LAE ($z=2.5$) samples using {\sc camb} \citep{lewis2000}. 
The modelling of the non-linear evolution of the power spectrum is based on a Galilean-invariant version of  renormalised perturbation theory \citep{crocce2006} dubbed gRPT, which will be presented in detail in
Crocce et al.\ 2021, in prep. \citep[see also the description in][]{eggemeier2020}. The gRPT model offers a good description of the power 
spectrum down to $k\leq0.25 h^{-1} \mathrm{Mpc}$ for a survey like BOSS \citep{sanchez2017}.

The bias model we use for the input power spectrum is from \citet{chan2012}, and it relates the 
galaxy overdensity $\delta_{\mathrm{g}}$ to the matter overdensity $\delta$ using
local bias parameters consisting of $b_{1}$ and $b_{2}$ and non-local bias parameters $\gamma_2$ and $\gamma_3^{-}$ as given in \citet{chan2012}.
The full expression of the power spectrum from this bias model is given in Appendix A of \citet{sanchez2017}. A review on perturbative bias is given in \citet{desjacques2018}

To generate input power spectra
 for the mocks, the $\gamma_{2}$ and $\gamma_{3}^{-}$ parameters are set following the local Lagrangian approximation \citep[see][]{fry1996, catelan1998, catelan2000, chan2012}. The local bias parameters we use for the LAEs and \oii galaxies are $b_{1}=2.5$ and $b_{1}=1.5$ respectively. The LAE bias we adopt is consistent with the $z \sim 2.5$ measurement 
of \citet{khostovan2019}, if we convert their power-law fits of the clustering to a bias via \citet{quadri2007}, which uses an expression from \citet{peebles1980}. The \oii galaxy bias is 
chosen to be consistent with previous work on HETDEX contamination \citep{grasshorn19}. For the LAE second order bias we use fitting functions of 
$b_{1}$ versus $b_{2}$ from \citet{lazeyras2016}, which they derive using the
separate universe approach of \citet{wagner2015}. We use the \citet{lazeyras2016} results at redshifts slightly higher than the maximum redshift they test, but they see no evidence of redshift dependence in their relations in the range they do test, $0<z<2$. The fitting function yields $b_{2}=0.986$ for the LAEs. We do not use the
same fitting function for the \oii galaxies, as it gives a negative power spectrum at 
scales important to the simulation. This is likely due to an insufficient 
number of terms in the expansion; correcting this issue would require higher order bias terms in the expansion. We therefore set $b_{2}=0$ for these galaxies since it gives a reasonable power spectrum. We do not model any dependence of the clustering of sources on luminosity or other galaxy properties as this should not impact our conclusions.

\subsection{Log-normal Simulations}
\label{sec:lognorm}
\begin{figure*}
\includegraphics[width=0.95\textwidth]{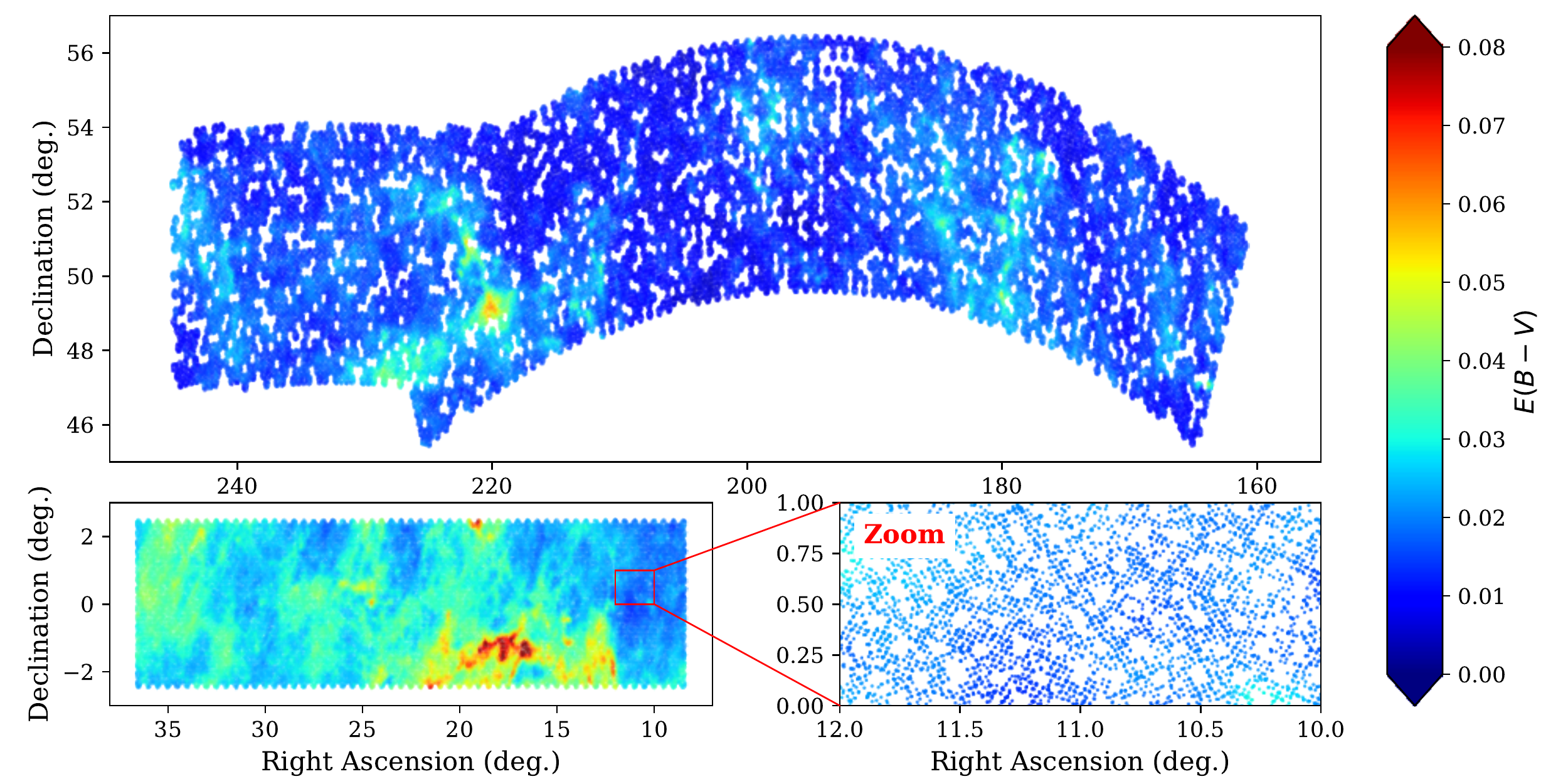}
\caption{A scatter plot of the sources in one of our mock HETDEX survey Spring fields (top) and Fall fields (bottom left), computed using an idealized focal plane containing 78 IFUs and a list of expected exposure set 
positions. The colour gives the reddening from Galactic extinction from the \citet{schlegel} dust maps. As the gaps between the IFUs are not visible on these two plots, we also show a zoom of the Fall field (bottom right).}\label{fig:footprint}
\end{figure*}
To generate mock catalogues with our desired power spectrum we use the log-normal simulation
code presented in \citet{agrawal2017}. A full explanation of the generation procedure is given in
the above paper, but we include a brief summary here. The code uses an input power spectrum $P^{G}(k)$ to generate a 3D Gaussian field on a grid in $k$-space, $G(\bf{k})$. It also generates random phases for each grid point and then carries out a Fourier transform to generate $G(\bf{x})$, a realisation of a Gaussian random field with the power spectrum $P^{G}(k)$. It then transforms this field to yield a field with a log-normal distribution, $\delta(\bf{x})$. The input power spectrum $P^{G}(k)$ is chosen in such a way that this resultant log-normal field will have the desired power spectrum $P(k)$. In this case our non-linear power spectrum is used for the matter density field, and our non-linear power spectrum with the added effects of bias is used for the galaxy density field. Each cell of the galaxy density field is randomly populated with galaxies. The number of galaxies assigned to a cell is drawn randomly from a Poisson distribution with a mean of $\bar{n}(1+\delta({\bf x}))V_{\mathrm{cell}}$, where $\bar{n}$ is the number density
of galaxies and $V_{\mathrm{cell}}$ is the cell's volume. The code also assigns a velocity to every cell using the linearised continuity equation in Fourier space on the simulated matter density field, using linear growth rates from {\sc camb} \citep{lewis2000}. Mock galaxies are then assigned the velocity of their cell.

The cell size we use in the simulations is 2.2~$h^{-1}$\,Mpc for our LAE mocks. For the mocks of the \oii galaxies, we use a minimum scale of 0.88~$h^{-1}$\,Mpc; the smaller size compensates for the fact \oii emitters are projected onto larger scales by their misclassification as LAEs. We expect resolution effects 
on scales to occur at least as small as twice the cell size, and we
will label this scale on our plots.

\subsection{Adding an Observer and the Angular Selection Function}\label{sec:alself}
To convert the simulated galaxies to a catalogue, we place an observer at an appropriate position in simulation coordinates and compute the right ascension,
declination and redshift to each mock galaxy from this observer's view point. The location of the observer, the simulation cube dimensions and the coordinate system are chosen in such a way to ensure the whole volume of a HETDEX field is contained within the simulation.  We assume the two widely separated Spring and Fall fields are independent, and we also assume the density fields of LAEs and \oii galaxies are independent (as in \citealt{addison2019} and \citealt{grasshorn19} we ignore the small, inferred correlations from gravitational lensing). We therefore simulate each population with separate log-normal simulations.

The line-of-sight (LOS) direction between the observer and every galaxy is computed, and each galaxy's velocity is projected onto the galaxy's LOS direction. These LOS velocities are used to apply the offsets to the galaxy's `observed' redshift, in order to model redshift space distortions (RSD). In these mocks we do not consider additional effects from the virial motions of galaxies within groups and clusters or from Ly~$\alpha$ radiative transfer \citep[see e.g.][]{behrens2018, byrohl2019, byrohl2020,gurung2019, gurung2020}. Also note that although this modelling uses linear-theory-derived velocities, the resultant power spectrum in redshift space is subject to the non-linear aspects of RSD which arise from the transformation of mock galaxies from cosmological to observed redshifts \citep{agrawal2017}.

We apply the angular footprints of the HETDEX fields
to the mock catalogues. The exposure set positions for the full survey are combined with the expected positions of the full 78 IFUs in the focal plane. Instead of using the actual mask for the data taken on the telescope we use idealized exposure set positions and assume a full focal plane from the start. We also assume 78 working units for this analysis, as there remains a goal to reach this number on the telescope. Having 74 working units is a more realistic expectation given the data taken at the time of writing (Gebhardt et al in prep). These small differences should not impact our conclusions on the decontamination.
Figure~\ref{fig:footprint} shows a mock catalogue with the angular selection function applied. The unusual shape of the Spring field is due to a decision (made in the first half of 2020) that the most efficient use of the telescope time is to extend the area rather than fill in missing regions from the originally planned footprint. This also explains the additional holes in the Spring footprint.

We use the masking software {\sc mangle} from \citet{swanson2008} and \citet{hamilton2004} to apply the survey footprint and also to generate a catalogue of random positions.
 These random positions are used to measure the clustering and we
 refer to them as the `random catalogue' or `randoms' hereafter. We also use {\sc mangle} to 
compute the area of the sky covered by fibers: 55.6~deg$^{2}$ in the Spring field and 27.2~deg$^{2}$ 
in the Fall field, and make use of a wrapper to {\sc mangle} called {\sc litemangle}\footnote[2]{\url{https://github.com/martinjameswhite/litemangle}}.
Although the footprint of HETDEX is unlikely to have any influence on our ability to discriminate LAEs from \oii galaxies, it does influence the error estimates we use 
to assess the size of systematic biases. 

These log-normal simulations are not true lightcone simulations like the ones used to probe contamination effects by \citet{massara2020}, or in the tomographic analysis of \citet{awan20}, since there is no evolution of the true power spectra along the redshift direction.
The focus of this paper, however, is to determine how the misclassification of \oii  emitters as LAEs produces a redshift dependent projection of the \oii galaxy density field, and how this redshift dependence, combined with redshift-dependent contamination fractions, affects clustering. These effects are included as we compute `observed' redshifts to all of our sources from a simulated observer's point of view. 

\subsection{LAE and \oii Properties}
\label{sec:lae_props}
To generate catalogues with realistic number densities and classification probabilities, we need to assign points in our mocks luminosities and equivalent widths (EWs). To assign LAE luminosities we use the 
Schechter  function fits to the $z=2.1$ and $z=3.1$ measured luminosity functions from \citet{gronwall2014}. These Schechter functions are parameterised by the characteristic luminosity, $L^{*}$,
the faint end slope, $\alpha$, and the number density coefficient, $\phi^{*}$. Similarly, we assume that the LAE EWs follow the exponential distributions found by \citet{gronwall2014}  
at those two redshifts \citep[see also equation 2 of ][]{leung}. 
The parameters for the \oii  luminosity and EW functions come from measurements in
four redshift bins between $z=0.1$ and $z=0.5$ by \citet{ciardullo2013}. 
At redshifts other than bin centers, we use the linearly interpolated or extrapolated values of all of the 
parameters. In Table~\ref{table:summary} we list the relevant parameters mentioned in this paragraph explicitly. The choice of 
luminosity and equivalent width functions are made to match the previous work on HETDEX source classification by \citet{leung}. We also use the approach of \citet{leung} to correct the measured luminosity functions for low EW LAEs (EW<20\,\AA), which were removed when the luminosity functions were estimated. We do not model any relationship between EW and luminosity as this level or realism is not needed for our work.

\subsection{Assigning Luminosities and the Radial Selection Function}\label{sec:rsel}
To apply the radial selection function to our mock and random catalogues we first begin by assigning
luminosities to our mock galaxies. A minimum luminosity is computed for each redshift assuming flux
limits much deeper than those of the survey: $6\times10^{-18}$~erg/s/cm$^2$ for LAEs and
$4\times10^{-18}$~erg/s/cm$^2$ for \oii galaxies. A maximum luminosity is computed as a large multiple of the minimum value, $L_{\mathrm{max}}=6000L_{\mathrm{min}}$.
To test if our choice of
$L_{\rm max}$ could affect results, larger values were 
tested. To avoid having to run the full simulation pipeline, we adopted a faster approach to test where we integrated products of the mean extinction, the luminosity function and
our completeness model in redshift slices up to an even higher $L_{\rm max}$. The number of sources predicted by our mocks and
this simple integration based technique agree to high precision ($<1$\% difference).

Between the two luminosity limits, random luminosities are drawn from our fiducial luminosity
functions (Table \ref{table:summary}). These luminosities are then translated to fluxes using the 
luminosity distance to the virtual observer. For the random catalogue distances are randomly chosen in a way that is uniform in volume, and luminosities and fluxes are drawn that are consistent with that distance.

Model flux limits are adopted using the
5$\sigma$ detection limit given in the HETDEX science requirements (and also presented in Hill et al, in prep), and 
are based on a typical sky spectrum along with the project's expectations for image quality 
and the efficiency of the whole telescope, spectrograph and detector system. Achieving
the number density of LAEs predicted by these flux limits is a target of HETDEX.

We divide the 5$\sigma$ flux limit at each mock galaxy's observer-frame wavelength by 5, and use that value as
the standard deviation of the Gaussian noise we add to the true flux of the emission line. The signal-to-noise 
(SNR) ratio of this noisy `observed' emission line is computed and the line is 
classified as `detected' 
in the simulation if its observed SNR exceeds 5. This results in sources being detected 50\% of the
time if their flux is exactly at the 5$\sigma$ limit; the completeness that corresponds to other fluxes can be computed 
by integrating a Gaussian and determining the area above the SNR cut. Carrying out these mock observations is more resource
intensive than simply applying the predicted $n(z)$ to the mocks, but it does have the benefit of including Eddington bias in
the emission line fluxes.  These fluxes are used to estimate the LAE/\oii galaxy probabilities (see section~\ref{sec:probs}). 

In addition to cuts in simulated SNR, another part of the radial selection function of \oii  emitters comes from their size.  As explained by \citet{leung}, at $z<0.05$ most \oii emitters will appear extended in imaging data and therefore easily distinguished from the LAE sample. We therefore do not include $z<0.05$ \oii emitters in our simulations. The argument that imaging
will be able to remove very nearby galaxies is also why
we do not consider the impact of other even longer rest-frame wavelength potential contaminants such as [\ion{O}{III}] emitters. We further 
discuss the possible impact of other contaminants on cosmology in an upcoming paper (Farrow et al, in prep.).

\subsection{On sky sensitivity variations and extinction}
\label{sec:ext}
The sensitivity and the flux errors of HETDEX vary from exposure set to exposure set, IFU to IFU and even fiber to fiber. These variations are likely to change how well we can classify LAEs as a function of sky position in the real survey. We do not model on-sky variations in the sensitivity of the survey, as in this work we focus on effects along the line of sight, which are likely much more important, since the clustering of the contaminants evolves quickly due to projection effects (see section~\ref{sec:proj}). 

Although we do not model the aforementioned on-sky sensitivity variations, we do add some sky-position dependent effects as we attenuate the fluxes and mock spectra (see section~\ref{sec:spectra}) by Galactic extinction. We model the sky-position dependence of Galactic extinction via the python library of \citet{green2018}, utilising the dust maps of \citet{schlegel}. To model the wavelength dependence we use
the {\sc extinction}\footnote[3]{\url{https://extinction.readthedocs.io/en/latest/}}
library with the \citet{fitzpatrick1999} function and the parameters advocated in the
Appendix of \citet{schlafly2010}. We confirm that our code can reproduce the extinction in the SDSS
bands \citep{doi2010} predicted by \citet{schlafly2011} using 
\citet{munari2005} stellar spectra to 2\% accuracy.  This is more than sufficient for our mock catalogues. We also add this extinction to the randoms, so it is accounted for when measuring the clustering.  

The Galactic dust reddening versus position
from \citet{schlegel} is indicated in Figure~\ref{fig:footprint}; the equatorial Fall field typically
has more Galactic extinction than the Spring field.
\begin{figure}
        \includegraphics[width=0.47\textwidth]{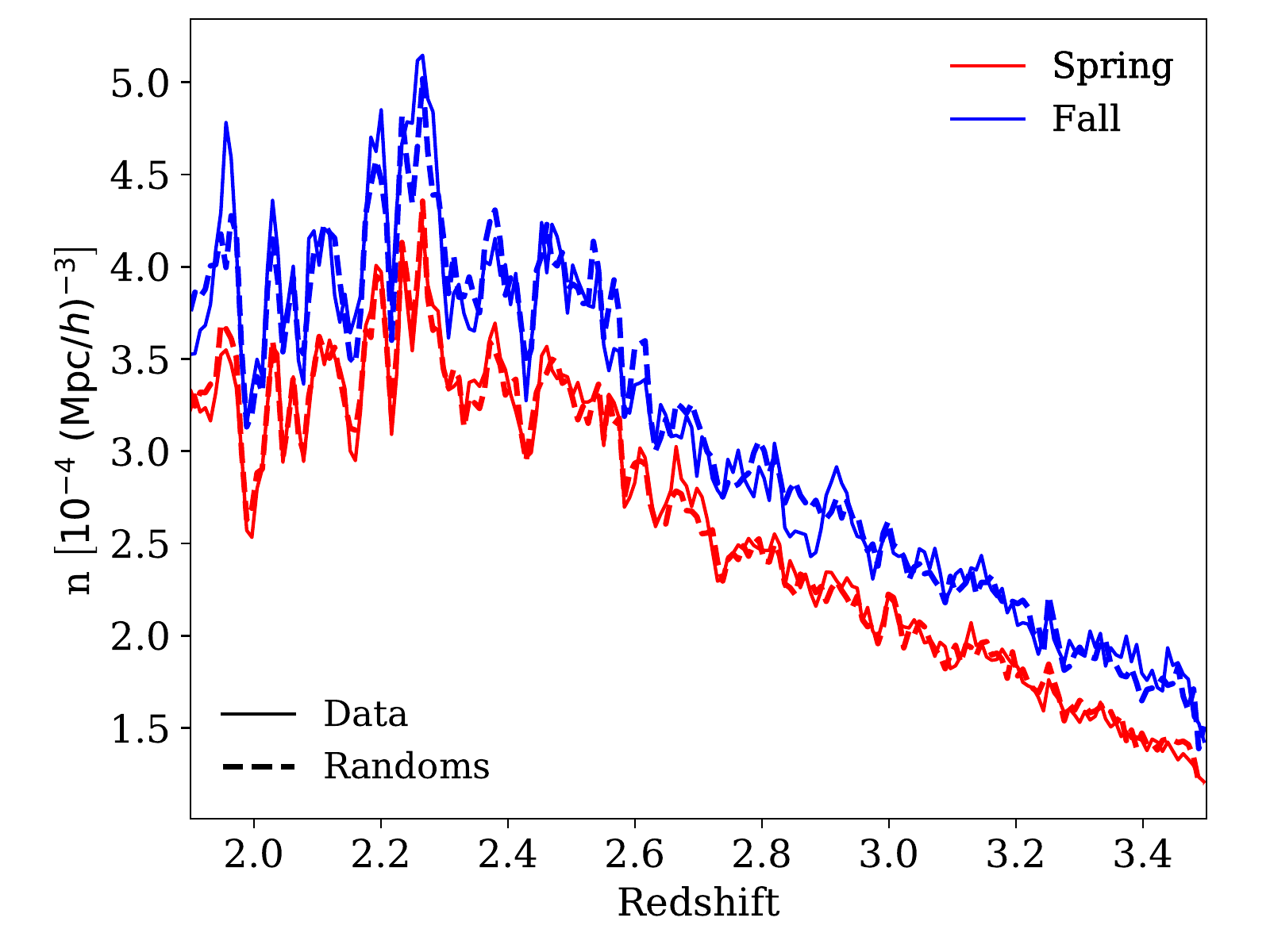}
        \caption{The number density of LAEs in one of our mock catalogues (solid lines) and random catalogues (dashed lines, normalized to the total number of mock sources) in the two HETDEX fields. The structure in the randoms is caused by the complex, wavelength dependent flux limits. The number density is computed assuming the sparsely sampled on-sky area of the two fields; the Fall field has higher number density as the fill-factor of the area is larger.}\label{fig:rsel}
\end{figure}

\subsection{Source density versus redshift}
In the following sections we continue our explanation of the mocks with how we assign mock continuum values and use them to classify galaxies as \lae\ or \oii. Before this, let us consider the number density of the mock catalogues without contamination. Our model of the selection function predicts about a million LAEs and 600,000 $z>0.05$ \oii emitters in the
full HETDEX survey. Figure~\ref{fig:rsel} shows the number density of detected 
emission line sources in one of our mock
catalogues (solid lines) and in our random catalogue (dashed lines) versus redshift. The plots are computed using the full volumes of the two fields.  
The most prominent troughs in the number density of the randoms are not due to noise, but the effect of sky lines, which propagate into the survey's sensitivity limit. The difference in the number density between the Spring and Fall fields is mostly caused by different sky filling factors (i.e. there are more gaps in the Spring field). This figure shows the effect of the complicated radial selection function on the detected number density.

\subsection{Mock LAE/\oii Galaxy Spectra}\label{sec:spectra}
In order to model the  separation of LAE and \oii emitters as accurately as possible, we generate mock spectra, which allow us to model the noise on the measured EWs more accurately. To do this we follow the approach of \citet{leung}. Details are available
in that paper, but summarising the method is helpful for future discussion. Equivalent widths are drawn from the distributions described in section~\ref{sec:lae_props},
with scale lengths as given in Table~\ref{table:summary}. A spectral slope is assigned to the line emitters,
based on $(g-r)$ colours in SDSS filters \citep{doi2010} randomly selected from a distribution that looks like the real data \citep[details in][]{leung}.
The line flux divided by the EW sets the amplitude of the mock spectra. Then absorption from the intergalactic medium is applied to the
mock spectra from the prescription in \citet{madau}, using code
adapted from \citet{leung} and \citet{acquaviva2011}.

We apply broad band filters to the mock spectra to simulate the imaging surveys
we intend to use to make estimates of the continuum flux density. In the Fall field we already have Dark Energy Camera \citep[DECam;][]{flaugher2015}
$r$-band survey data from the \textit{Spitzer}/HETDEX Exploratory Large Area survey \citep[SHELA;][]{papovich2016, wold2019} and the Dark Energy Survey \citep[DES;][]{abbott2018}, so we apply the DECam $r$-filter \citep{abbott2018}. In the Spring field we have complete coverage with 
Hyper-Suprime Cam (HSC) data in the $r$-band, so we apply the HSC filter \citep{kawanomoto2018}. We use the Python library {\sc speclite}\footnote[4]{Note we use the older `DECam 2014' filters, see the {\sc speclite} website for details 
(\url{https://speclite.readthedocs.io/en/latest/index.html}). Using the older filter curves should not impact our conclusions.} to supply the filter response functions. Noise is added to the mock magnitude
measurements, using a rough estimate derived by dividing the 5$\sigma$ flux limits of the SHELA
survey by five. The 5$\sigma$ sky-aperture magnitude limits of SHELA were determined by \citet{wold2019}, and we take the
mean of the four different fields in this work, $r=24.6$, converting to flux via \citet{oke83}. For simplicity, we use the noise based off of SHELA for the whole survey, which in some areas is actually covered by DES or HSC. This simplification has some impact on the precision of the assigned probabilities, but should not affect the conclusions of our work. Also, early analysis suggests the HSC
data is significantly deeper than the SHELA data, so in the Spring field this is a conservative approach. The noisy magnitude measurements are combined with
the noisy line flux measurements to make a noisy estimate of the equivalent width, $EW_{\mathrm{obs}}$. 

A few subtleties are worth mentioning here. Firstly, although all
of the noise we add is Gaussian, the distribution of $EW_{\mathrm{obs}}$ can be realistically non-Gaussian due to taking the inverse of the noisy continuum estimates.
Secondly, note we make the assumption in our mock EW observations that the continuum is flat across the $r$-band and the spectral range of HETDEX\null. In real data more sophisticated techniques could be used, but here we again decide to be conservative and make the most simple mock measurements from our spectra. 
Finally, note that for the broad bands we use, which are to the red of Ly~$\alpha$, applying IGM absorption makes no difference to the results, but we include it in the model for possible future work.

Also following the \citet{leung} approach, we add other expected emission lines to the spectra of \oii galaxies (namely [\ion{Ne}{III}] $\lambda 3869$~\AA, H$\beta$ $\lambda 4959$~\AA, [\ion{O}{III}]  $\lambda 4949$~\AA\ and [\ion{O}{III}]  $\lambda 5007$~\AA)
using fixed line ratios for one fifth solar abundance \citep[][and references therein]{anders2003}. We also add appropriate Gaussian noise to these lines, following the same wavelength dependent noise prediction used for Ly~$\alpha$. These other emission lines can also be used to identify \oii emitters in the regions of redshift where they are within the spectral range of HETDEX\null. We use this method to generate 1000 realistic mock HETDEX catalogues.

\subsection{A modified method to assign probabilities}\label{sec:probs}

To split the mocks into `observed' LAE and \oii samples, we assign each mock source a probability
of being an LAE, based on its `observed' properties. To generate these probabilities, we reformulate the Bayesian method of separating the two classes that was presented in \citet{leung}. We begin by
presenting a conceptually different way to formulate the problem, that results in a set of more easily evaluated
equations. We use the same set of inputs as in \citet{leung}, except for the source colour as it is unclear whether
we will have deep multiband imaging over the whole HETDEX field. We then consider a small $n$-dimensional box in the parameter space of
EW, flux, wavelength, and the flux of other non-[\ion{O}{II}]/LAE emission lines. Assuming the primary emission-line can only be \oii $\lambda 3727$ or Ly~$\alpha$ the probability of the source being an LAE is
\begin{equation}
        \label{eqn:prob}
        P_{\lae} = \frac{N_{\lae}}{N_{\lae} + N_{\oii}},
\end{equation}
where $N_{\lae}$ and $N_{\oii}$ represent the 
number of LAE and \oii emitters, respectively, in the box defined in the space of parameters used for the discrimination. We want this box to
be a fixed size in observed coordinates. If we choose a fractional interval of $\pm\delta$ in observed flux ($f$), equivalent
width, ($w$) and wavelength ($\lambda$) this corresponds to
\begin{align}
        &(1 \pm \delta)L = (1 \pm \delta)f \cdot 4\pi d_{\mathrm{L}}^2,\\
        &(1 \pm \delta)w = (1 \pm \delta)w_{\mathrm{obs}}/(1 + z), \\
        &(1 \pm \delta)(z+1)- 1 = (1 \pm \delta)\lambda/\lambda_{\mathrm{line}} - 1
\end{align}
where $d_{\mathrm{L}}$ is the luminosity distance. We can now express the number in terms of integrals over the luminosity function, $\Phi(L/L_{*}, z)\,\mathrm{d}L/L^*$, the equivalent width distribution $W(w, z)$ and a Gaussian, $G(f_{\mathrm{obs}} - f_{\mathrm{exp}},\sigma_{\mathrm{line}})$, with mean $f_{\mathrm{exp}}$ and    dispersion  $\sigma_{\mathrm{line}}$.  This last term expresses the difference between the noisy measured flux and the expected
flux, $f_{\mathrm{exp}}$, of a non-\oii emission line (i.e., [\ion{Ne}{II}], [\ion{O}{II}] etc.), in terms of the uncertainty in the measurement,  $\sigma_{\mathrm{line}}$.  This term is the product over all of the other emission lines that are expected, given the wavelength of detection and assuming the galaxy is an \oii emitter. The expression for the expected number of LAEs or \oii galaxies is then
\begin{align}
\begin{split}
        N  = \int_{(z+1)(1 - \delta) - 1}^{(z + 1)(1 + \delta) - 1}\frac{\mathrm{d}V}{\mathrm{d}z}\mathrm{d}z'
             \int_{L(1 - \delta)}^{L(1 + \delta)}\Phi(L'/L_{*}, z)\mathrm{d}(L'/L_{*})\\
             \times \prod_{i=\mathrm{[OIII],[H\beta],...}}\int_{f_{\mathrm{obs},i}(1 - \delta)}^{f_{\mathrm{obs},i}(1 + \delta)}\,G(f' - f_{\mathrm{exp},i}, \sigma_{i}) \mathrm{d}f'\\
             \times \int_{w(1 - \delta)}^{w(1 + \delta)}W(w', z)\mathrm{d}w'.
\end{split}
\end{align}
This equation is very similar to equation (19) of \citet{leung}, except here we do not normalise by the number density of the emission
line sources at the redshift under consideration. Moreover, \citet{leung} chose a fixed size
value for $\delta$; we set $\delta$ to be infinitesimally small as then we can drop the integrals. The number in an
infinitesimally sized box becomes
\begin{multline}
        N  = \frac{\mathrm{d}V}{\mathrm{d}z}2\delta (z + 1) \cdot
             \Phi(L'/L_{*}, z)2\delta L/L_{*} \cdot W(w', z)2\delta w \\
             \times \prod_{i=\mathrm{[OIII],[H\beta],...}}G(f_ {\mathrm{obs},i} - f_{\mathrm{exp},i}, \sigma_{i})2\delta f_{\mathrm{obs},i}.
\end{multline}
Then, using equation (\ref{eqn:prob}), substituting $1+z$ with the ratio of observed to assumed rest wavelength, and cancelling the $2\delta$ and $\lambda$ terms, the
expression for the LAE probability becomes
\begin{equation}
        \label{eqn:probstart}
        P_{\lae} = \frac{\tilde{N}_{\lae}}{\tilde{N}_{\lae} + \tilde{N}_{\oii}},\\
\end{equation}
with
\begin{multline}
\label{eqn:probend}
        \tilde{N}_{x} = \Lambda_x \frac{\mathrm{d}V}{\mathrm{d}z}
                                    \Phi(L_{x/L_{*,x}}, z_{x}) \frac{L_{x}}{L_{*,x}} W_x(w_{x}, z_{x}) w_{x}\\
                                    \times \prod_{i=\mathrm{[OIII],[H\beta],...}}G(f_{\mathrm{obs},i} - f_{\mathrm{exp},i}^x, \sigma_{i}) f_{\mathrm{obs},i},
\end{multline}
where $x$ labels whether the relevant functions and measurements are for LAEs or \oii galaxies, $\Lambda_{\lae}=1$ and
$\Lambda_{\oii}=\lambda_{\lae}/\lambda_{\oii}$. For LAEs, the expected flux at the wavelength of
other emission lines is $f_{\mathrm{exp},i}^{\lae}=0$, while for \oii  emitters, this value is equal to the
relative line ratio for each line, $R_{i}$, multiplied by the observed \oii flux, i.e.,
$f_{\mathrm{exp},i}^{\oii}=R_{i}f_{\mathrm{obs},\oii}$. In these simulations we evaluate equation~(\ref{eqn:probend}) using the true underlying input luminosity and equivalent width distributions, the input line ratios, and cosmology used in the survey. We also use our mock observed measurements when computing the probabilities, which adds noise similar to real data. Future HETDEX papers will carry out more extensive tests and assessments of LAE classification approaches (Davis et al, in prep).

Our library to produce these probabilities, and also an implementation of the \citet{leung} method, has been integrated into the rest of the HETDEX source classification code, and is also available online\footnote[5]{\url{https://github.com/djfarrow/hetdex-line-classification}}. The authors of \citet{leung} 
provided us with their original code, which we use as a reference (and for some sections reproduce directly) in our implementation. This is also true for parts of the HETDEX simulation pipeline. 

\subsection{The mock observed LAE and \oii samples}
To generate samples of contaminated LAEs and \oii emitters from the mocks, we 
classify all sources with $P_{\lae}>0.5$ as LAE and all
other sources as \oii galaxies. Despite the fact that these probabilities do not
account for the noise on the $EW_{\mathrm{obs}}$ or on the LAE/\oii line
flux, this simple cut produces an LAE sample where only 1.3 per cent of the 
sources are misclassified \oii emitters and 4.4 per cent of the  observed \oii catalogue are LAEs. This is actually better than the target LAE sample contamination fraction of 2 per cent, but our classifier is better than what is obtainable for real data, as it assumes we know the properties of the input LAE and \oii populations perfectly. To consider a pessimistic scenario we also 
split the samples using a less conservative cut 
of $P_{\lae}>0.15$, which produces a purer \oii sample (contamination fraction of 1.7 per cent), but a greater number of contaminants in the LAE sample (5.1 per cent).  It might seem surprising that the $P_{\lae}>0.15$ cut still gives a relatively small fraction of contaminants, but it is important to 
realise the $P_{\lae}$ values assigned to individual \oii emitters are skewed towards zero, as for most sources, the classification is nearly unambiguous.  In the rest of the paper we will refer to the \emph{high contamination} sample as that for $P_{\lae}>0.15$ and the \emph{low contamination} sample for $P_{\lae}>0.5$. These two samples bracket the expected 2 per cent contamination of HETDEX.
\begin{figure}
    \centering
    \includegraphics[width=0.45\textwidth]{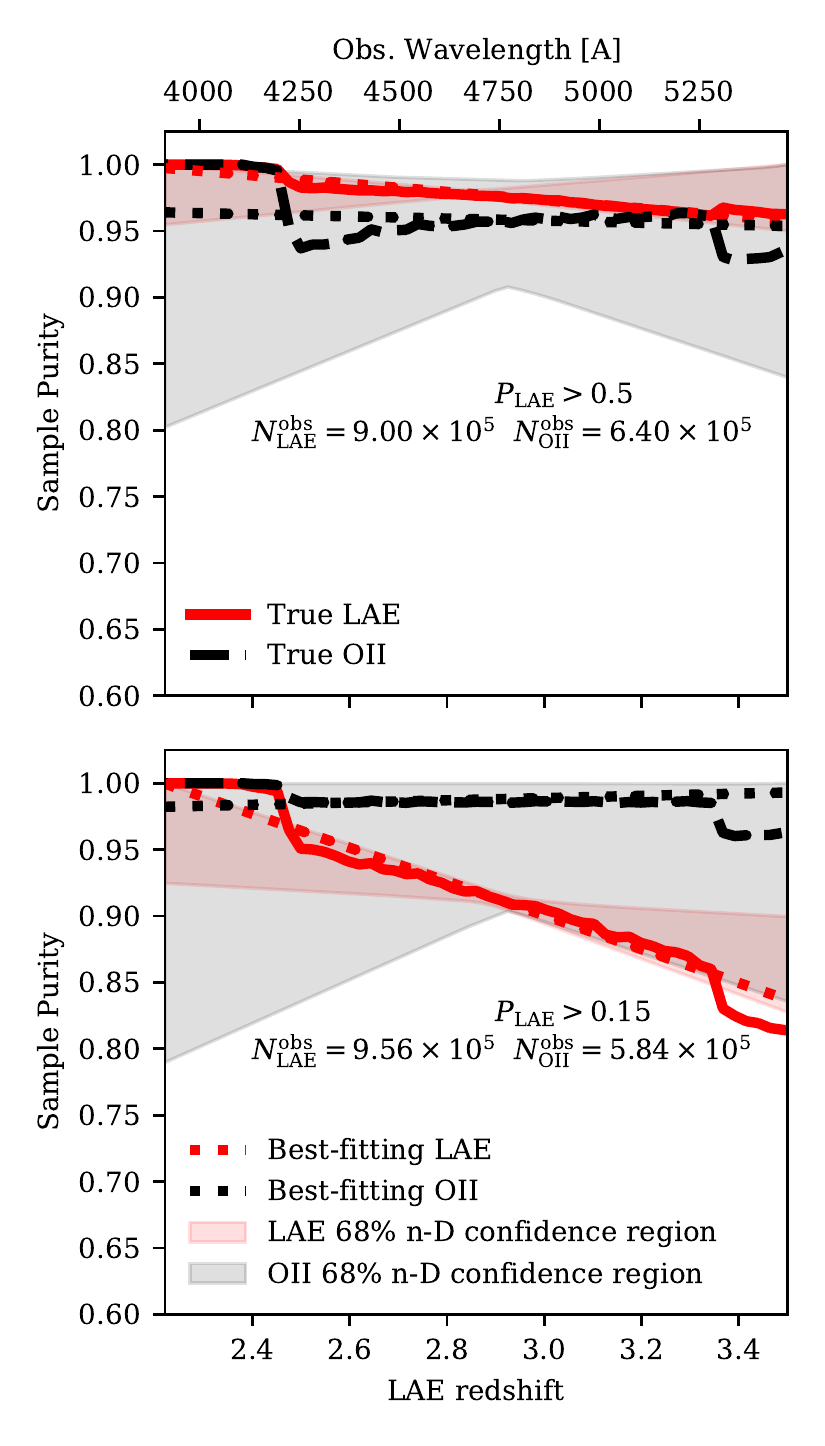}
    \caption{The purity of the mock `observed' LAE (solid red lines) 
             and \oii (dashed black lines) galaxy catalogues for the P$_{\lae}$=0.5 cut (top) and the P$_{\lae}$=0.15 cut (bottom). We only show the observed wavelength range where \oii emitters are included in the simulation. The sharp drops occur 
                 where important emission lines redshift out of the HETDEX spectral range, specifically [OIII] $\lambda$5007 [OIII] $\lambda$4949, H$\beta$ and [NeIII] at $z=2.35, 2.38, 2.45$ and $3.34$ respectively. The inset numbers show the total number of sources in the full HETDEX redshift range in each of the samples (including interlopers) for the given cuts. The dotted lines show the best-fitting contamination values from our linear model of LAE and \oii purity, which has two parameters per galaxy type: $f(z_{\rm low})$ and $f(z_{\rm high})$. The shaded regions show the maximum and minimum purity values in the 68\% confidence region (see section~\ref{sec:purityc}).}
    \label{fig:purityz}
\end{figure}

In order to create a random catalogue that correctly follows the redshift distribution of the data samples, we also compute LAE probabilities for the random catalogue and apply the same probability cuts. If we used 
random catalogues without contamination the different redshift distribution of the randoms versus that inferred for the observed samples would 
cause a huge systematic bias.

The predicted sample purity, defined as the number of correctly classified sources in a sample divided by the total size of the sample, is shown in Figure~\ref{fig:purityz}. The lower redshift limit of this plot corresponds to our minimum redshift for \oii emitters ($z=0.05$). Although our simulations make
the simplifying assumption of perfect knowledge of the true distribution of \oii and LAE properties, we can still see many features expected for LAE/\oii classifiers. As the 
observed emission line wavelength
increases, the volume of space inhabited by \oii emitters grows faster than
that of the LAEs, causing a decrease in the purity of the LAE sample. The
large, sudden decreases in the purity correspond to where emission lines useful in identifying a source as an \oii emitter are redshifted out of the HETDEX spectral range. 
Although the full, high contamination LAE sample has an interloper fraction of 5.1 per cent, when the sample is split by redshift the contamination can be as large as around 17 per cent in the highest redshift bins.

\section{Correlation Functions}\label{sec:corrf}

\subsection{Measuring the Clustering}

The correlation functions of the mock catalogues are measured on a two-dimensional grid of the galaxy and/or random pair separation, $s$,
and the cosine of the angle between the pair separation vector and the line of sight, $\mu$. We  use the 
estimator introduced by \citet{landy1993}, modified for cross-correlation functions by \citet{blake2006},
\begin{equation}
\xi(s, \mu) = \frac{DD_{\rm c}(s, \mu) - D_{\rm c}R(s, \mu) - DR_{\rm c}(s, \mu) + RR_{\rm c}(s, \mu)}{RR_{\rm c}(s, \mu)},
\end{equation}
where 
c indicates which of the objects in the pair is an \oii emitter and 
$DD_{\rm c}(s, \mu)$, $D_{\rm c}R(s, \mu)$, $DR_{\rm c}(s, \mu)$ and $RR_{\rm c}(s, \mu)$ are the binned counts of pairs of LAEs and \oii galaxies, \oii galaxies and LAE randoms, LAEs and \oii randoms, and LAE randoms and \oii randoms, respectively. 
The auto-correlation functions are estimated with the usual \citet{landy1993} estimator.
We compute the line of sight direction to each pair of galaxies as the vector between the observer and the mid point of the separation vector of the pair.
When measuring the correlation function, we use random LAE and/or \oii catalogues at least 13 times larger than the data catalogue, to decrease shot noise from the randoms. We measure the auto-correlation functions and the cross correlation functions  assuming Ly~$\alpha$ derived redshifts for both the LAE and \oii catalogues, except when measuring the $\oii$ clustering to use with equation~\ref{eqn:proj}, where we use the $\oii$ derived redshifts. 

The 2D correlation functions are integrated in $\mu$, weighted with
the appropriate Legendre polynomials, to yield measurements of the first three even multipoles, $\xi_{\ell}(s)$, following the standard 
method \citep[e.g.][]{sanchez2017}. 
The covariance matrix is estimated from the measured multipoles also using the standard approach, i.e.,
\begin{equation}
C_{\ell\ell'}(s_a, s_b) =\frac{1}{N_{\mathrm{mk}} - 1} \sum\limits_{i=0}^{N_{\mathrm{mk}}}(\xi_{\ell}(s_a) - \bar{\xi}_{\ell}(s_{a}))(\xi_{\ell'}(s_{b}) - \bar{\xi}_{\ell'}(s_{b})),
\end{equation}
where $C_{\ell \ell'}(s_a, s_b)$ is the covariance between multipoles $\ell$ and $\ell'$, for measurement bins $s_{a}$ and $s_{b}$. The index $i$ runs over
the number of mock catalogues, $N_{\mathrm{mk}}=1000$. The quantities with bars, e.g.,  $\bar{\xi}_{\ell'}(x_{b})$, are the mean values from all of the mock catalogues.

The simulated Fall and Spring fields have different average flux limits due to different values of the Galactic extinction. Normally if the fields have significantly different average flux
limits they would be biased differently and need to be analysed separately. In our simulations however all the LAE sources have the same correlation function; we therefore combine the two fields by computing weighted sums of the multipoles and covariances following equations~(8) and (9) of \citet{white2011}. 

\subsection{Projected \oii clustering}\label{sec:proj}
The \oii contaminants in the LAE sample
are assigned redshifts assuming the rest-frame wavelength of Ly~$\alpha$, and vice-versa for the
LAE contaminants in the \oii sample. The relation between the source redshift assuming the emission line is \oii $\lambda 3727$ rather than Ly~$\alpha$ is simply
given by
\begin{equation}
\label{eqn:zconv}
z_{\oii} = (1 + z_{\lae})\frac{\lambda_{\lae}}{\lambda_{\oii}} - 1.
\end{equation}
As noted in \citet{lidz2016} the misclassification has an effect very analogous to the Alcock-Paczynski test \citep[hereafter AP; ][]{alcock79}, in that the three dimensional positions inferred
from the position and redshift of the sources are distorted. Following \citet{pullen2016,leung} and the earlier similar derivation from \citet{visbal2010}
while adopting a slightly different notation, we can relate the true separation of a pair of \oii emitters, in directions parallel, $s'_{\parallel}$, and perpendicular, $s'_{\perp}$, to the line of sight, to the separation projected into LAE coordinates ($s_{\perp}$, $s_{\parallel}$) by misclassification with \begin{equation}
s'_{\perp} = s_{\perp}c_{\perp}, \qquad s'_{\parallel} = s_{\parallel}c_{\parallel}
\end{equation}
with
\begin{equation}
    c_{\parallel}(z_{\lae}) = \frac{\lambda_{\lae}} {\lambda_{\oii}}\frac{H(z_{\lae})}{H(z_{\oii})},\qquad
    c_{\perp}(z_{\lae}) = \frac{D_\mathrm{M}(z_{\oii})}{D_\mathrm{M}(z_{\lae})},
\label{eqn:disto}
\end{equation}
where $H(z)$ is the Hubble parameter and $D_\mathrm{M}(z)$ is the comoving angular diameter distance to $z$. This is given by $D_{\mathrm{M}}(z)=(1+z)D_{\mathrm{A}}(z)$, where $D_{\mathrm{A}}(z)$ is the angular diameter distance. Given these distortion parameters, the correlation function  can be written as
\begin{equation}
\label{eqn:proj}
\xi_{\oii}^{\proj}(s, \mu, z_{\lae}) = \xi_{\oii}(sq(\mu), \mu c_{\parallel}(z_{\lae})/q(\mu)),
\end{equation}
where we do not explicitly show the dependence of $q$ on $z_{\lae}$ to shorten the notation \citep[for the expression for the power spectrum see e.g.,][]{pullen2016,leung,grasshorn19}. Equation~(\ref{eqn:proj}) assumes all the evolution of the projected \oii clustering is caused by projection effects, as is the case in our simulations. It should be possible in future work to extend this methodology to also include intrinsic evolution of the \oii correlation function. The value of $q$ is given by \citet{ballinger96} \citep[see also e.g., equation 9 of][]{pullen2016}
\begin{equation}
\label{eqn:q}
q(\mu) = [c_{\parallel}^2(z_{\lae})(\mu)^2 + c_{\perp}^2(z_{\lae})(1 - (\mu)^2)]^{1/2}.
\end{equation} 
The equations describing the clustering of LAEs misclassified as \oii galaxies are the same but
with the inverse of the distortion parameters, i.e. $c_{\parallel}^{-1}$ and $c_{\perp}^{-1}$. 
As the distortion parameters are an approximation 
of a more complicated effect, we carry out tests in appendix~\ref{sec:appendix} of the distortion parameters compared to a brute force approach. This appendix also presents an additional test of the methodology we present in section~\ref{sec:lf_decon}.
\begin{figure}
    \centering
    \includegraphics[width=0.45\textwidth]{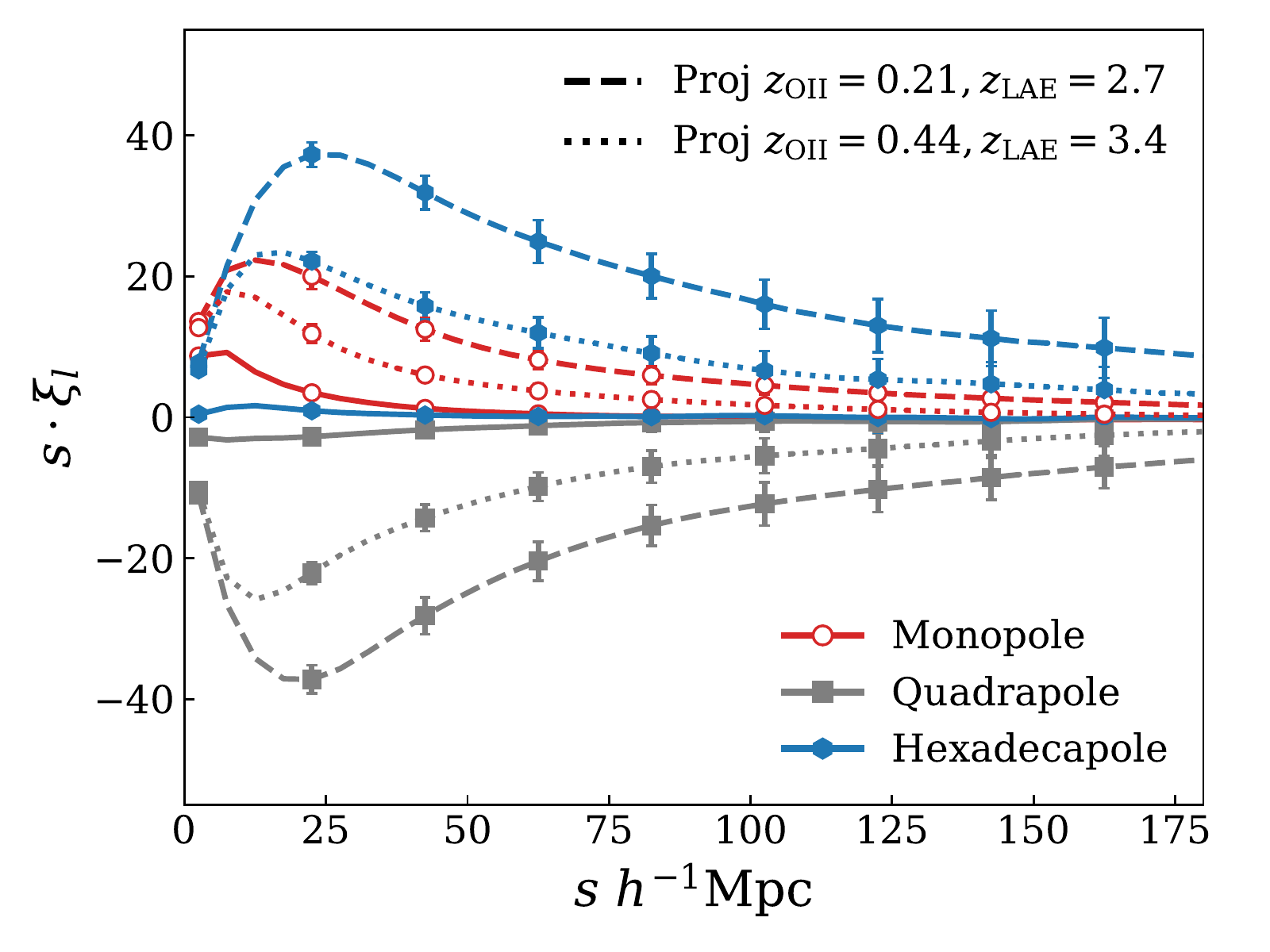}
    \caption{The solid lines show mean of the \oii galaxy correlation function multipoles measured from 199 of our mock catalogues, along with error bars expected from a single realisation of HETDEX\null. The dotted and dashed lines show the multipoles when distorted by a projection to different LAE redshifts, as indicated in the legend. This projection occurs due to LAE/\oii misclassification and we model it using equation~(\ref{eqn:proj}). For visual clarity, only every 4th data point and error bar is marked, and the correlation functions have been multiplied by the separation, $s$.}
    \label{fig:oii_proj}
\end{figure}

The redshift dependence of the distortions causes 
the clustering of the \oii contaminants to evolve with (Ly~$\alpha$ based) redshift. To illustrate these effects we show in Figure~\ref{fig:oii_proj} the mean correlation function measured from 199 pure mock \oii catalogues, along with the same measurements projected onto two different Ly~$\alpha$ redshifts. To predict the projected measurements, we use equation~(\ref{eqn:proj}), linearly interpolating over the measured \oii correlation function for $\xi_{\oii}$. The solid lines show multipoles from the samples analysed with the
\oii redshifts, the negative quadrapole is evidence of the Kaiser effect \citep{kaiser87}, an effect of the peculiar velocities of galaxies 
falling into over-densities. The dashed line shows the predictions of projecting from the \oii redshift at $z_{\oii}=0.21$ to the misclassified LAE redshift of $z_{\lae}=2.7$. We see the projection  causes a clear increase in the monopole for all but the smallest separations under consideration. 
We also see the quadrupole becomes much more negative, which is a result of the projected correlation function appearing very elongated along the direction transverse to the line of sight. The impact of this on the multipoles is
much larger than for the Kaiser effect. In Fourier space the elongation looks like a compression along the direction transverse to the line of sight \citep[see e.g., Figure 3 of][]{grasshorn19}. We also note an increase in the hexadecapole. 

The dotted lines in Figure~\ref{fig:oii_proj} show the predicted multipoles of $z_{\oii}=0.44$ \oii emitters that are misclassified as $z_{\lae}=3.4$ LAEs. We see similar trends to the lower redshift projection, but the amplitude of the distorted multipoles is lower. This decrease is driven by $c_{\perp}$ becoming closer to unity, and the distortion transverse to the line of sight is much larger than the distortion in the parallel direction, $c_{\parallel}$ \citep[e.g.][]{grasshorn19}. We will return to modelling these signals over a redshift range in section~\ref{sec:lf_decon}.

At this point we highlight the fact that 
we always use the true cosmology when computing the parameters for the projection. As highlighted by \citet{addison2019}, if we want to make predictions for the projected functions in the real data, we need to be aware of the 
additional
uncertainty from not knowing the actual cosmology. We discuss this again at the end of the paper. 

\section{Decontamination Methods}
\label{sec:deconmed}

\subsection{Simple decontamination ignoring redshift dependencies}
As mentioned, in this paper we develop a new method to deal with the redshift dependence of the contamination. We start by slightly modifying equation (12) of \citet{awan20} to use the multipoles of the two-dimensional correlation function instead of the angular clustering, giving
\begin{equation}
\label{eqn:con_old}
[\xi^{\obs}_{\mathrm{\ell, aa}}(s), \xi^{\obs}_{\mathrm{\ell, ab}}(s), \xi^{\obs}_{\mathrm{\ell, bb}}(s)]^{\rm T} = \mathbfss{D}_{\rm s}[\xi^{\true}_{\mathrm{\ell, aa}}(s), \xi^{\true}_{\mathrm{\ell, ab}}(s), \xi^{\true}_{\mathrm{\ell, bb}}(s)]^{\rm T},
\end{equation}
 where $\ell$ indicates the multipole, `a' and `b' indicate the two possible samples (in our case LAEs and \oii galaxies), the `true' and `obs' superscripts indicate the pure and contaminated correlation functions and $\mathbfss{D}_{\rm s}$ is the contamination matrix. 
The matrix of
\citet{awan20} compactly expresses the important equations for contamination, which have also been presented in other literature \citep[e.g.][]{pullen2016, leung, grasshorn19, addison2019}. The matrix contains contributions from the fractions of each type of galaxy that were correctly classified (i.e., the purity), labelled $f_{\mathrm{aa}}$, and $f_{\mathrm{bb}}$, and the fractions that were misclassified, $f_{\rm ab}$ and $f_{\rm ba}$. 
In \citet{awan20} this matrix is given as
\begin{equation}
\label{eqn:autodecon}
\mathbfss{D}_{\rm s}=
\begin{pmatrix}
f^2_{\mathrm{aa}} & 2f_{\mathrm{aa}}f_{\mathrm{ab}} & f^2_{\mathrm{ab}} \\
f_{\mathrm{aa}}f_{\mathrm{ba}} & f_{\mathrm{aa}}f_{\mathrm{bb}} + f_{\mathrm{ab}}f_{\mathrm{ba}} & f_{\mathrm{ab}}f_{\mathrm{bb}} \\
f^2_{\mathrm{ba}} & 2f_{\mathrm{bb}}f_{\mathrm{ba}} & f^2_{\mathrm{bb}} \\
\end{pmatrix},
\end{equation}
where the contamination fractions can be computed from the purity via $f_{\mathrm{ba}}=1 - f_{\mathrm{bb}}$ and $f_{\mathrm{ab}} = 1 - f_{\mathrm{aa}}$. 
To be more specific to the case of HETDEX, we relabel $f_{\rm aa}$ as $f_{\lae}$ and $f_{\rm bb}$ as $f_{\oii}$.
Also following \citet{awan20}, the decontaminated estimates of the auto and cross correlation functions can then be given by applying the matrix inverse to a vector of the observed functions, i.e.,
\begin{equation}
\label{eqn:decon_old}
[\xi_{\mathrm{\ell, aa}}^{\rm est}(s), \xi_{\mathrm{\ell, ab}}^{\rm est}(s), \xi_{\mathrm{\ell, bb}}^{\rm est}(s)]^{\rm T} = \mathbfss{D}_{\rm s}^{-1}[\xi^{\obs}_{\mathrm{\ell, aa}}(s), \xi^{\obs}_{\mathrm{\ell, ab}}(s), \xi^{\obs}_{\mathrm{\ell, bb}}(s)]^{\rm T}.
\end{equation}
The superscript `est' indicates the decontaminated estimates of the correlation function. Again, for the specific case of HETDEX $\xi_{\ell, \rm aa}(s)$, $\xi_{\ell, \rm bb}(s)$ and $\xi_{\ell, \rm ab}(s)$ are the auto-correlation functions of the \lae\ sample, $\xi_{\ell, \lae}(s)$, the \oii sample, $\xi_{\ell,\oii}(s)$, and the cross-correlation $\xi_{\ell,\lae \times \oii}(s)$ respectively.
Once we have estimates of the auto-correlation functions, we can make an estimate for the contribution of the contamination to the observed cross-correlation signal, $\xi^{\mathrm{pred, obs}}_{\mathrm{\ell, \lae \times \oii}}(s)$,  using equation~(\ref{eqn:con_old}) resulting in
\begin{equation}
\label{eqn:resid_old}
\begin{aligned}
\xi^{\mathrm{pred, obs}}_{\mathrm{\ell, \lae \times \oii}}(s) &= f_{\lae}(1-f_{\oii})\xi_{\ell, \lae}^{\rm est}(s)\\ + &f_{\oii}(1-f_{\lae})\xi_{\ell, \oii}^{\rm est}(s).
\end{aligned}
\end{equation}
This can be related to the full decontaminated cross-correlation from equations~(\ref{eqn:con_old}) and (\ref{eqn:autodecon}) via
\begin{equation}
\label{eqn:resid_to_decon}
\xi^{\mathrm{est}}_{\mathrm{\ell, \lae \times \oii}}(s) = \frac{\xi^{\mathrm{obs}}_{\mathrm{\ell, \lae \times \oii}}(s) - \xi^{\mathrm{pred, obs}}_{\mathrm{\ell, \lae \times \oii}}(s)}{f_{\oii}f_{\lae} + (1-f_{\oii})(1 - f_{\lae})}.
\end{equation}
We will label this approach `simple decontamination' and differentiate it from our new approach of `lightcone decontamination'. We note that \citet{awan20} developed this method for angular clustering 
in tomographic redshift bins. They do not claim that the method will work for our scenario, which has rapidly evolving projected OII contamination within the redshift bins considered. However, we present it in its unmodified form as a demonstration of what might happen if one does not take additional steps to deal with this rapid evolution.

\subsection{Lightcone based decontamination} \label{sec:lf_decon}
When we apply the matrix of \citet{awan20} to HETDEX, we make the assumption that the clustering of the galaxies classified as \oii emitters is the same as the clustering of \oii interlopers in the LAE sample with some fixed scaling for contamination. However, this
may not be the case, as the shape of the volume number density versus redshift, $n(z)$, of the interlopers will not match that of the \oii sample when the purity has a redshift dependence. To give a hypothetical example, consider most of the \oii emitters being at the high redshift end of the range. If that were the case, the projected clustering of the \oii sample would have distortion parameters appropriate for high redshifts. If all of the misclassifications occurred at low redshift however, then the interlopers would have low-redshift distortion parameters.

The idea then, is to use something like the 
decontamination matrix of \citet{awan20}, but instead of using the observed clustering of the \oii emitters, we apply a prediction for the clustering of contaminants that is consistent with the redshift dependence of the interloper number density, $n^{\mathrm{inter}}(z)$. To make a prediction for the expected interloper clustering in a redshift range, we refer to the work of \citet{yamamoto1999} and \citet{suto2000}.
They approximate the correlation function between two galaxies as
the correlation function at the mid-point between them \citep[equation 19 of][]{yamamoto1999}. This results in a fairly intuitive expression that approximates the
observed correlation function for galaxies in the redshift range $z_{\mathrm{min}}$ to $z_{\mathrm{max}}$ as an integral of the redshift-dependent correlation function weighted by the square of the number density as a function of redshift, i.e.
\begin{equation}
        \label{eqn:lc}
        \xi_{\ell}^{LC}(s) = \frac{\int^{z_{\mathrm{max}}}_{z_{\mathrm{min}}} \mathrm{d}z\frac{\mathrm{d}V}{\mathrm{d}z} n(z)^2 \xi_{\ell}(s;z)}{\int^{z_{\mathrm{max}}}_{z_{\mathrm{min}}} \mathrm{d}z\frac{\mathrm{d}V}{\mathrm{d}z} n(z)^2 }.
\end{equation}
This differs slightly from equation (18) of \citet{suto2000} in that we use the observed number density of objects, not the true comoving number density in real space, so that the terms related to the selection function and the AP distortion are unneeded. Equation~(\ref{eqn:lc}) also assumes that $n(z)$ does not change much over the separations under consideration, $s<180\,h^{-1}$\,Mpc, and the redshift evolution of $\xi_{\ell}(s;z)$ is slow enough to be unimportant over those same scales.
This is an approximation, as there are certainly 
redshifts over which the projected \oii clustering changes rapidly.  But as we will see, the simplification works reasonably well for our simulations. For surveys whose properties differ from those of mock HETDEX, it would be prudent to test the technique with tailored simulations.   

To continue, we define a function to carry out the lightcone (i.e., redshift) integral, $\mathcal{F}(x, y)$, as
\begin{equation}
\label{eqn:lc_func}
\mathcal{F}(x(z), y(z)) = \frac{\int^{z_{\mathrm{max}}}_{z_{\mathrm{min}}} \mathrm{d}z\frac{\mathrm{d}V}{\mathrm{d}z} x(z)^2y(z)}{\int^{z_{\mathrm{max}}}_{z_{\mathrm{min}}} \mathrm{d}z\frac{\mathrm{d}V}{\mathrm{d}z} x(z)^2}.
\end{equation}
Given equation~(\ref{eqn:lc_func}), and using equations~(\ref{eqn:con_old}) and (\ref{eqn:autodecon}) with equation~(\ref{eqn:lc}), we get the following expression for the prediction of the observed auto-correlation function for LAEs with redshift $z$ and purity $f(z)$,
\begin{equation}
\begin{aligned}
\label{eqn:lcobs}
\xi^{\LC, \obs}(s, \mu) &=  \mathcal{F}[n_{\lae}(z),\, f^2(z)\xi_{\lae}^{\true}(s, \mu, z)] \\&   + \mathcal{F}[n_{\lae}(z),\, (1 - f(z))^2\xi_{\oii}^{\proj}(s, \mu, z)]\\& + 
2\mathcal{F}[n_{\lae}(z),\, f(z)(1 - f(z))\xi_{\mathrm{\lae \times \oii}}^{\true,\proj}(s, \mu, z)
].
\end{aligned}
\end{equation}
The subscripts on $n(z)$ indicate which observed sample redshift versus volume number density should be used. To be closer to the numerical implementation we replaced the multipoles of equation~(\ref{eqn:con_old}) with the 2D correlation function; this makes no practical difference as the decontamination has no $\mu$ dependence so the order of decontamination and converting the measurements to multipoles is unimportant. The number densities are for the total "observed" samples with contaminants. Here we use the fact that the (LAE) redshift distribution of the \oii interlopers is given by $n_{\oii}^{\mathrm{inter.}}(z)=(1 - f(z))n_{\lae}(z)$. We explicitly include the redshift dependence of the purity parameters, to differentiate them from their redshift independent versions: $f_{\lae}$ and $f_{\oii}$. For brevity we also drop the \lae\ subscript from the redshift dependent purity parameter in this section. Since the integral of the number density gives the total number of objects, the redshift dependent and independent types of purity parameter are related by the lightcone integral, i.e. for the \lae\ sample
\begin{equation}
    f_{\lae} = \mathcal{F}(\sqrt{n_{\lae}(z)}, f(z)),
\end{equation}
and the equivalent for the \oii sample.

Samples of emission line galaxies with nearby rest-frame wavelengths, such as H$\beta$ and \oiii, will have a non-zero cross-correlation due to large-scale structure, and even distant samples like our LAE and \oii galaxies will have a slightly non-zero cross-correlation due to cosmic magnification of the background galaxies by the foreground galaxies. As mentioned, our simulations do not include such magnification, as it is a very small signal.
We therefore will now assume that the true cross-correlation between the \oii and LAE samples is zero. This limits the method to scenarios, like HETDEX, where the contaminants are not correlated with the main sample.
However, future work on surveys with sample-contaminant cross-correlations could still use equation~(\ref{eqn:lcobs}) in 
an approach that tries to forward-model the relevant auto- and cross- correlations. 

Given that the cross-correlation is zero, to find our estimate for  $\xi_{\lae}^{\true}(s, \mu, z)$
we now make the assumption that the LAE clustering does not evolve with redshift, which allows us to take it out of the lightcone integral and we are left with
\begin{equation}
\label{eqn:trueest}
\begin{aligned}
    \xi_{\lae}^{\est}(s, \mu) = \mathcal{F}[ n_\lae(z),\, f^2(z)]^{-1}\{\xi^{\LC, \obs}(s, \mu)\\ - \mathcal{F}[n_\lae(z),\, (1 - f(z))^2\xi_{\oii}^{\proj}(s, \mu, z)]\}.
\end{aligned}
\end{equation}
The integrals over redshift are then all carried out numerically and an estimate of the true correlation function, $\xi_{\lae}^{\est}$, can be made. Even when the assumption that $\xi_{\lae}^{\true}(s, \mu, z)$ does not evolve over the whole survey is unreasonable, this approach can be utilized to estimate the observed correlation function in bins of redshift over which that evolution is expected to be small enough that this average provides a meaningful observable. Additionally, we discuss plans to relax this assumption in section~\ref{sec:conc}.

Given our estimate of $\xi_{\lae}^{\est}(s, \mu)$, we can use the cross-correlation term of  equations~(\ref{eqn:con_old}) and (\ref{eqn:autodecon})
to predict the observed cross-correlation measured with LAE  redshifts as follows:  \begin{equation}
\label{eqn:deconx2}
\begin{aligned}
 &\xi^{\mathrm{LC,obs}}_{\mathrm{\lae \times \oii}}(s, \mu) = \\
 &\mathcal{F}[\{n_{\lae}(z)n^{\mathrm{proj}}_{\oii}(z)\}^{0.5},\,f(z)(1-f_{\oii}(z))\xi_{\lae}^{\est}(s, \mu)] + \\
   & \mathcal{F}[\{n_{\lae}(z)n^{\mathrm{proj}}_{\oii}(z)\}^{0.5},\,f_{\oii}(z)(1-f(z))\xi_{\oii}^{\proj}(s, \mu, z)] + \\
& \mathcal{F}[\{n_{\lae}(z)n^{\mathrm{proj}}_{\oii}(z)\}^{0.5}, \\
& \,\,\,\,\,\,\{f(z)f_{\oii}(z)+(1-f(z))(1-f_{\oii}(z))\}\xi_{\mathrm{\lae \times \oii}}^{\true,\proj}(s, \mu, z)],
\end{aligned}
\end{equation}
where we have restored the cross-correlation term, $\xi_{\mathrm{\lae \times \oii}}^{\true,\proj}(s, \mu, z)$. Here $n^{\mathrm{proj}}_{\oii}(z)$ is the redshift distribution of \oii emitters projected into LAE redshifts. This latter 
relation can easily be measured by computing redshifts assuming Ly~$\alpha$ for the galaxies in the \oii sample. We cannot estimate the redshift dependent cross-correlation
by simply rearranging this expression, but if the 
cross-correlation is expected to be non-zero a forward modelling approach could be used. Here we forward-model the expected cross-correlation signal, using the fact $\xi_{\mathrm{\lae \times \oii}}^{\true,\proj}(s, \mu, z)=0$ in HETDEX, which gives
\begin{equation}
\label{eqn:deconx}
\begin{aligned}
&\xi^{\mathrm{pred, obs}}_{\mathrm{\lae \times \oii}}(s, \mu) = \\
 &\mathcal{F}[\{n_{\lae}(z)n^{\mathrm{proj}}_{\oii}(z)\}^{0.5},\,f(z)(1-f_{\oii}(z))\xi_{\lae}^{\rm est}(s, \mu)] + \\
   & \mathcal{F}[\{n_{\lae}(z)n^{\mathrm{proj}}_{\oii}(z)\}^{0.5},\,f_{\oii}(z)(1-f(z))\xi_{\oii}^{\proj}(s, \mu, z)].
\end{aligned}
\end{equation}
This equation is the lightcone version of equation~(\ref{eqn:resid_old}). The use of the lightcone equations requires a model of the 
projected \oii clustering. As in section~\ref{sec:proj}, we interpolate over the measured clustering of the \oii sample, and then apply a redshift dependent projection via equation (\ref{eqn:proj}). To make a fairer comparison of decontamination techniques, we only interpolate over the measurement of the \oii clustering for each single realisation of the catalogue, and we use the observed \oii catalogue, not the pure one. As the observed \oii clustering has contamination, we experimented with an iterative process where we first decontaminate the \oii clustering with the projected LAE correlation function, and then use the decontaminated \oii clustering to decontaminate the LAE clustering. We find the first and second iterations give almost identical results, so we stop after two iterations and use the resultant \oii clustering to decontaminate the LAE measurement.

\section{Results}
\label{sec:results}
The differences between the multipoles of the auto-correlation function for the (de)contaminated and the pure cases (measured from the corresponding pure LAE catalogues) for the full HETDEX redshift range ($1.9<z<3.5$) are given in the left column of Figure~\ref{fig:full_range_xi}. 
The upper and lower panels in Figure~\ref{fig:full_range_xi} give the two $P_{\lae}$ cuts under consideration. The points are the mean of the 1000 mocks. Each measurement has been divided by the statistical error expected for the HETDEX survey, i.e., the square-root of the diagonal of the covariance matrix derived from the mocks. As we have 1000 mock catalogues, 
the errors on our mean measurement are much smaller than the statistical error on a single HETDEX mock. In the following, when we refer to $\sigma$, we specifically mean the statistical error on a single realisation.

The right column shows for the same redshift range, the residual difference between the observed cross correlations of the \oii and LAE samples and 
the predicted cross correlation from the 
contamination, from equation~(\ref{eqn:resid_old}) for the simple method and equation~(\ref{eqn:deconx}) for the lightcone method. The differences are divided by
the statistical errors. Recall from equation~(\ref{eqn:resid_to_decon}) that for the simple method the decontaminated cross-correlation is related to the
residual we plot, $\xi^{\mathrm{obs}}_{\mathrm{\ell, \lae \times \oii}}(s) - \xi^{\mathrm{pred, obs}}_{\mathrm{\ell, \lae \times \oii}}(s)$,  by a constant factor. As we divide by
the statistical error this means the residuals we plot for the simple decontamination method are also the statistical significance (ignoring the diagonal terms of the covariance) of the 
spurious cross-correlation signal that remains in the simple decontaminated multipoles. We chose to frame the discussion of the simple method in terms of the residuals, in order to make comparisons to the lightcone approach easier.

In the following subsections we study the impact of contamination on the raw (section~\ref{sec:raw}), the simple decontaminated (section~\ref{sec:simple_decon}) and lightcone decontaminated (section~\ref{sec:lc_decon}) multipoles. The final subsection considers the scenario where the redshift range with no modelled \oii emitters ($z<0.05$) is cut out of the catalogue (section~\ref{sec:red_zrange}).

\subsection{Raw clustering multipoles}
\label{sec:raw}
\begin{figure*}
    \centering
    \begin{tabular}{cc}
      \includegraphics[width=0.45\textwidth]{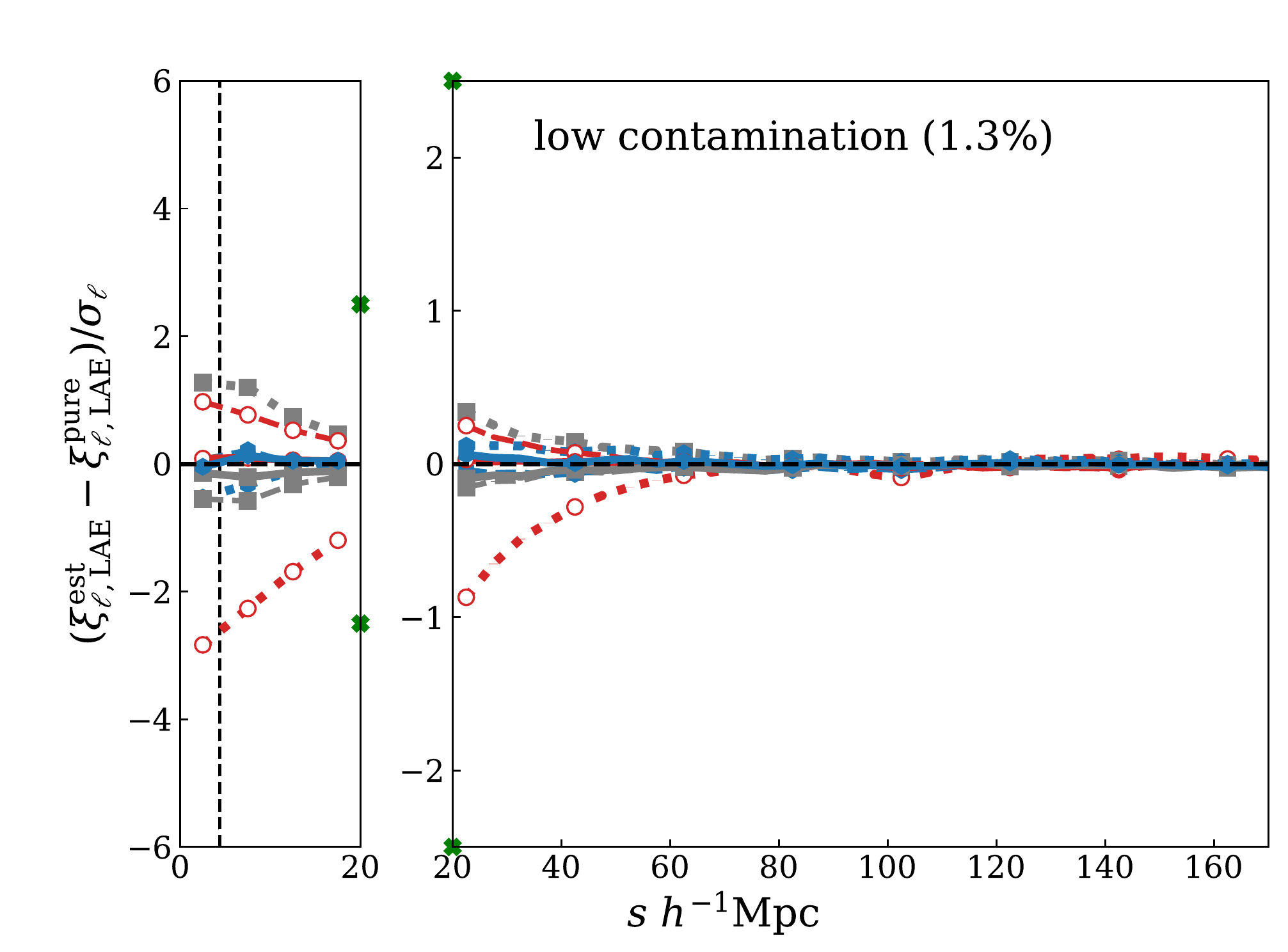} & \includegraphics[width=0.45\textwidth]{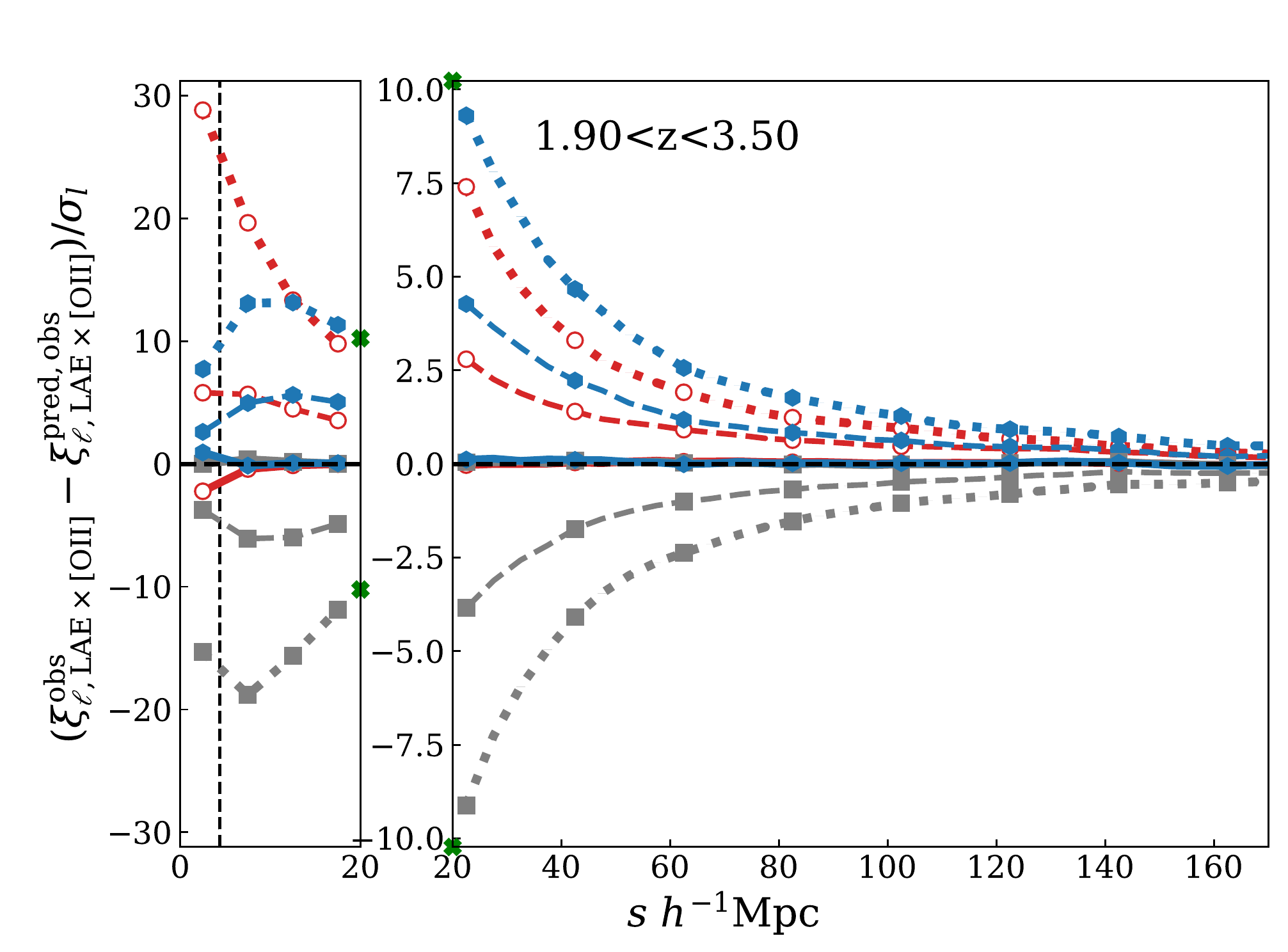} \\
      \includegraphics[width=0.45\textwidth]{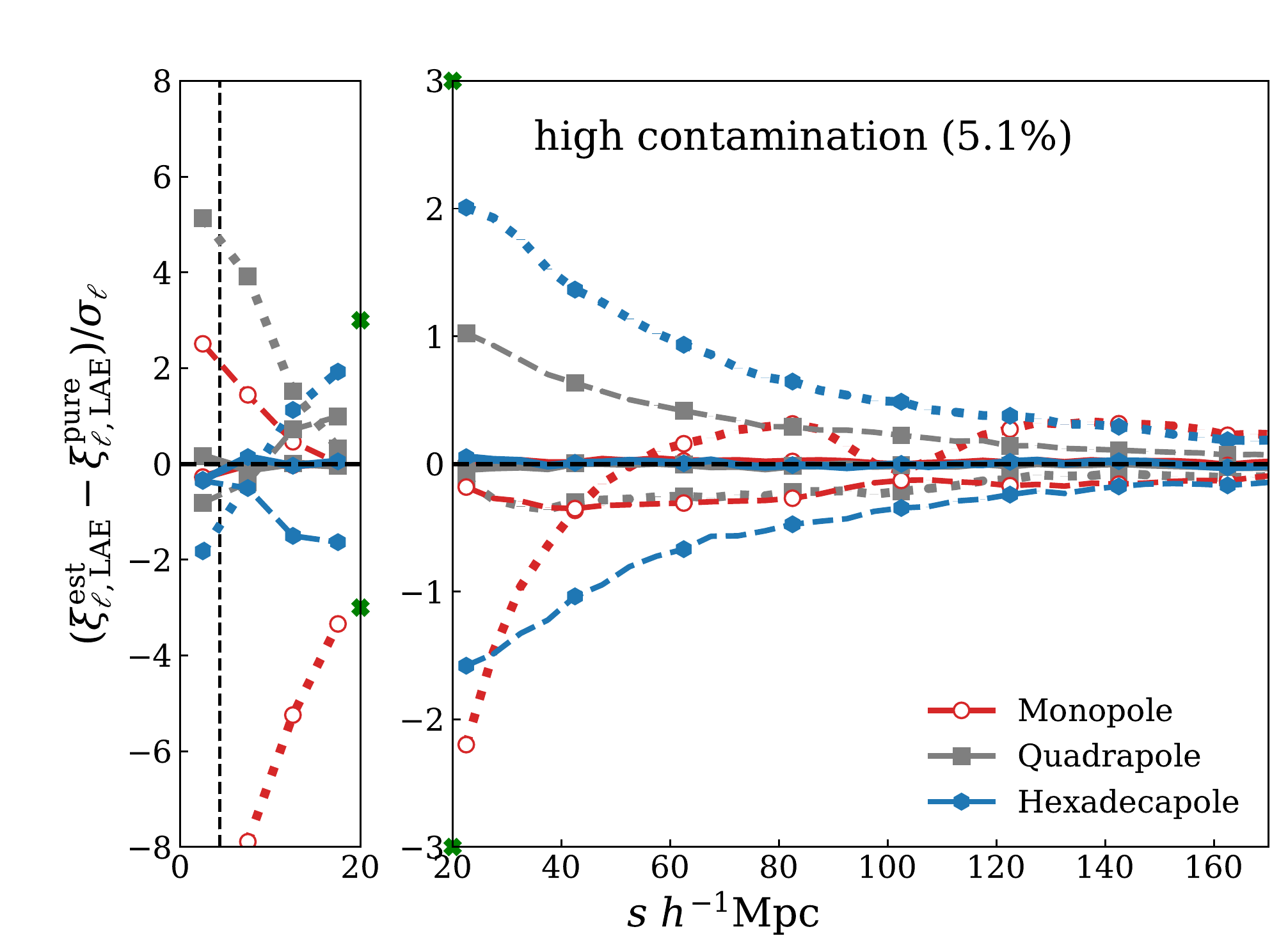} & \includegraphics[width=0.45\textwidth]{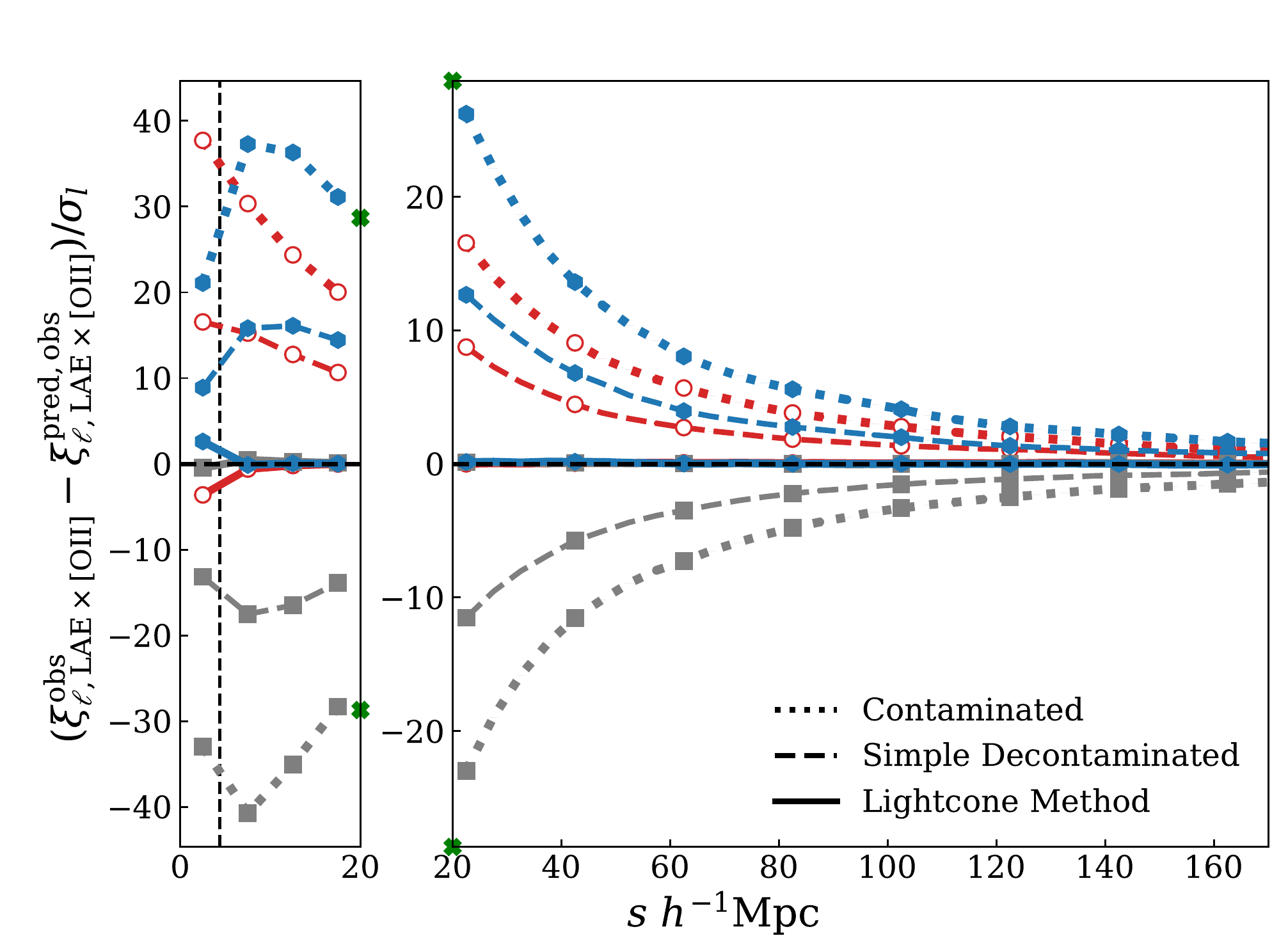} \\
    \end{tabular}
    \caption{The left column shows the average difference between the LAE correlation function from 1000 mock LAE catalogues containing $\oii$ contamination and the pure catalogues. 
    Each panel is split at $s=20 h^{-1}$Mpc to enable a better vertical and horizontal dynamic range, green crosses indicate the range of the right panel on the axes of the left. In the right panel only every forth data point is marked with a symbol for clarity.
    The right column shows the residual signal left when subtracting the predicted cross-correlation from contamination from the 
    measured cross-correlation functions, averaged over the 1000 mock \oii and LAE catalogues. Different approaches to dealing with 
    contamination are shown: dotted lines display results with no corrections, dashed lines show the results of a decontamination procedure which ignores the redshift dependence of the interlopers (`simple decontaminated') and the solid lines show results from our new method that accounts for redshift dependencies (`lightcone decontaminated'). If the contamination is fully accounted for, all the differences plotted here should be zero. All results are divided by the error on a single realisation, to give a rough estimate of the statistical significance of any unaccounted for contamination in either method.
    The colours indicate the mono-(red), quadru- (gray) and hexadeca- (blue) poles  of the correlation function. The different rows show results from the different LAE probability cuts used to define samples: 1.3 per cent contamination (top) and 5.1 per cent 
    contamination (bottom), see Figure~\ref{fig:purityz} for the purity versus redshift of these samples. The dashed, black
    vertical line shows twice the cell size of the LAE simulation box.}
    \label{fig:full_range_xi}
\end{figure*}

It is clear from the dotted lines Figure~\ref{fig:full_range_xi}, which show the results from the raw measurements of the contaminated mocks, that interloper contamination modifies the correlation signals.  This has been previously seen or predicted many times in the literature \citep[e.g.][]{leung, awan20, grasshorn19, addison2019, massara2020}. 
In the low contamination LAE sample, the offset in the auto-correlation 
is generally $<1\sigma$ but 
reaches a maximum difference of $\sim2.5\sigma$ at scales down to twice the cell size of the log-normal simulations (indicated by the vertical dashed line). In the high contamination sample, where 5.1 per cent of the LAE catalogue are \oii galaxies, many of the auto-correlation function hexadecapole measurements are systematically high by up to 2$\sigma$ at separations $>20~h^{-1}$Mpc. At 
smaller scales,
where the effect of the contamination appears greater, the raw auto-correlation function  
multipole measurements can be biased by several $\sigma$.

The signal of contamination in the cross-correlation function is even stronger. We can see from the right column of Figure~\ref{fig:full_range_xi} that we expect the signal to be clearly detected in the HETDEX survey, even at low levels of contamination. 
The cross-correlation function, which is zero for pure samples, shows a positive monopole and hexadecapole, and a negative quadrupole. The negative quadrupole implies that the two-dimensional cross-correlation function appears elongated transverse to the line of sight. This could be explained by the fact that the cross-correlation is dominated by the strong clustering signal of \oii galaxies \citep{grasshorn19}, which, as mentioned, appears elongated due to projection effects.
\subsection{Simple Decontaminated measurements}
\label{sec:simple_decon}
In this section, we use equation~(\ref{eqn:decon_old}) to decontaminate the galaxy samples while ignoring redshift dependencies in the interloper fraction.  We take the required contamination and purity factors, i.e. the components of $\mathbfss{D}_{\rm s}$, from the numbers of LAEs and \oii emitters in the mock catalogues averaged over the realisations; in real data, some procedure will be needed to measure the contamination and purity.  We cover this scenario in section~\ref{sec:fit}.

The dashed lines in Figure~\ref{fig:full_range_xi}, labelled `simple decontaminated', show the results. In the low contamination LAE sample (1.3 per cent \oii galaxies), the 
simple method results in all of the multipoles having no significant systematic bias ($<1.0\sigma$) all the way down to the resolution limit of the catalogue. The decontaminated measurements are a modest improvement 
over the raw measurements. In the high LAE sample contamination  case (5.1 per cent \oii emitters) the decontamination improves the monopole, but at small scales the monopole still has a bias approaching $\sim 2 \sigma$. Additionally the hexadecapole is biased up to $\sim 1.5\sigma$ low after the decontamination --- a bias in the opposite direction from what was seen in the raw data. 

Subtracting the predicted contribution to the cross-correlation from contamination decreases 
the observed cross-correlation. However, the correction is too small, and very significant ($>5\sigma$ over a range of scales) cross-correlations still remain in the high contamination case. The low contamination sample also shows residual cross-correlation signals that increase relative to the statistical errors. As smaller separations are considered, these signals increase, causing what can be as great as a $\sim5\sigma$ spurious signal. As mentioned, the residuals we plot are related to the full
decontaminated cross-correlation from the simple method via a constant factor (equation~\ref{eqn:resid_to_decon}), meaning they also show the significance of 
spurious cross-correlations left after the full decontamination process. Although a strong cross-correlation signal would not be expected to directly affect cosmological parameters derived from HETDEX, a non-zero cross-correlation after simple decontamination does suggest a failure of the 
modelling which can cause indirect effects. For example, the methods presented in \citet{grasshorn19} and \citet{addison2019} use the cross correlation signal to determine the purity and contamination of the LAE and \oii samples.  If the lightcone effects are ignored, then inferred values of the contamination will become artificially high in order to force the decontaminated cross correlation toward zero. This in turn would impact the contamination and purity values used to decontaminate the auto-correlation, causing additional bias in their measurement. 
We will demonstrate this in section~\ref{sec:fit}.
\begin{figure*}
    \centering
    \begin{tabular}{cc}
      \includegraphics[width=0.45\textwidth]{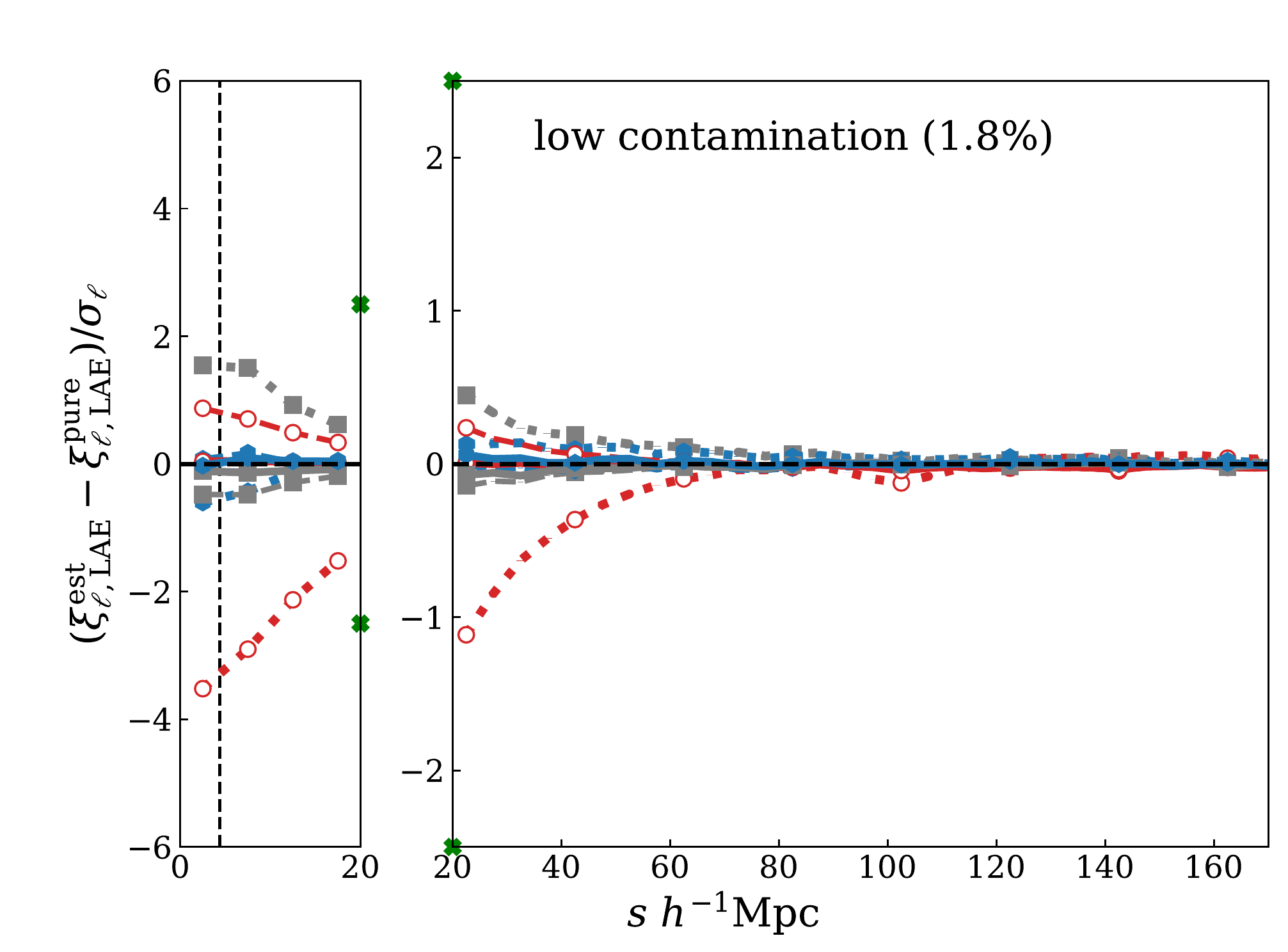} & \includegraphics[width=0.45\textwidth]{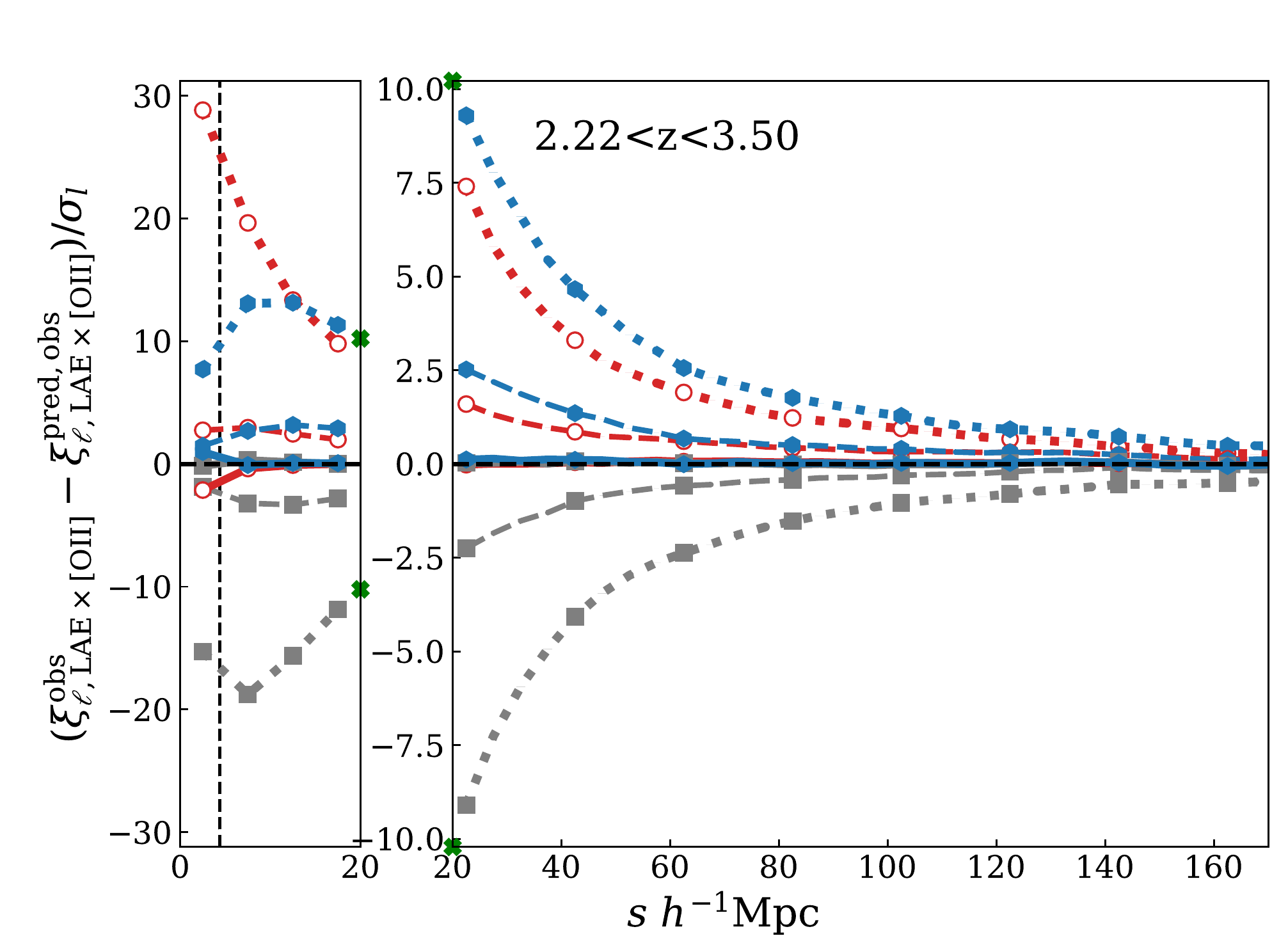} \\
      \includegraphics[width=0.45\textwidth]{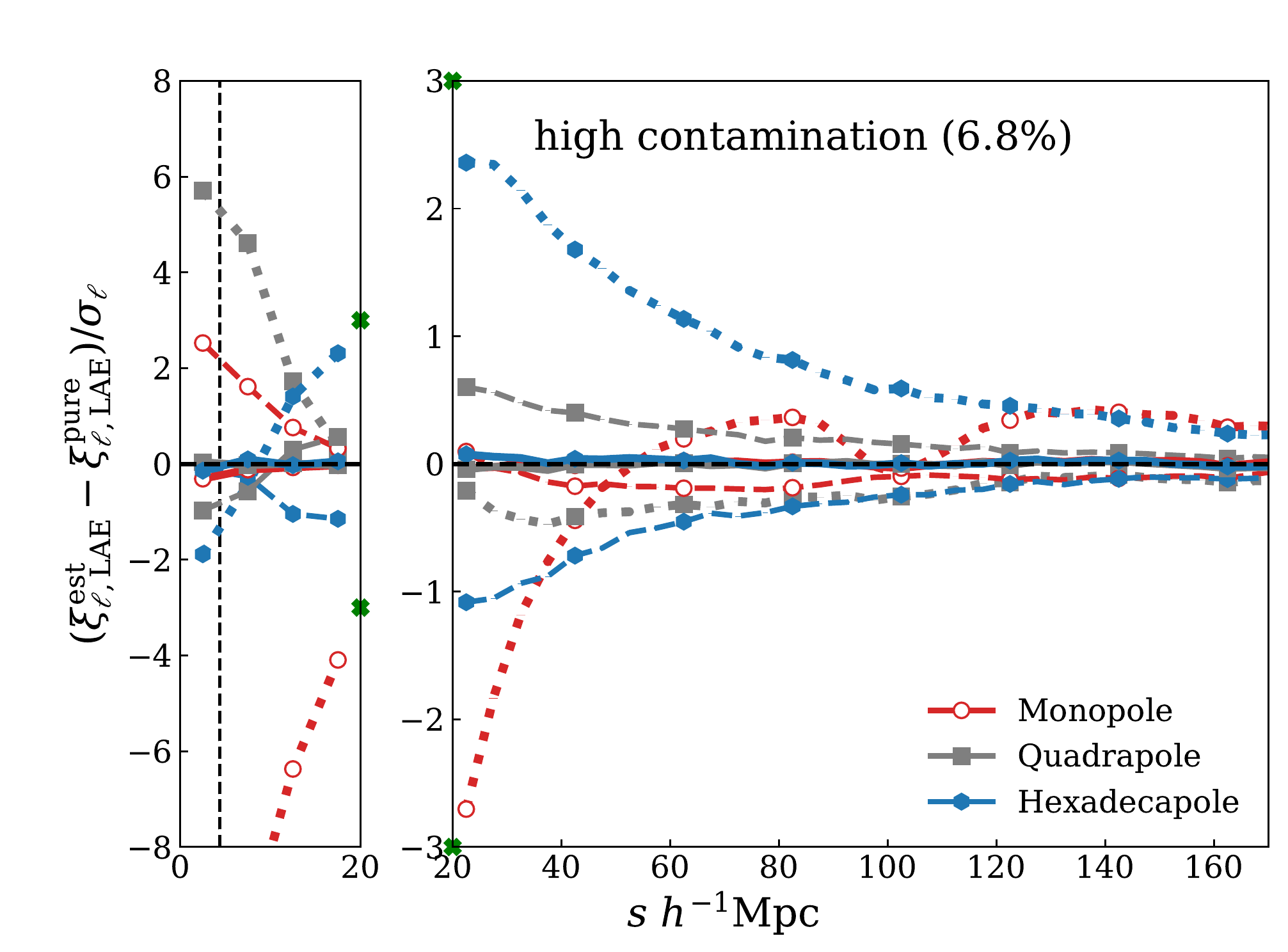} & \includegraphics[width=0.45\textwidth]{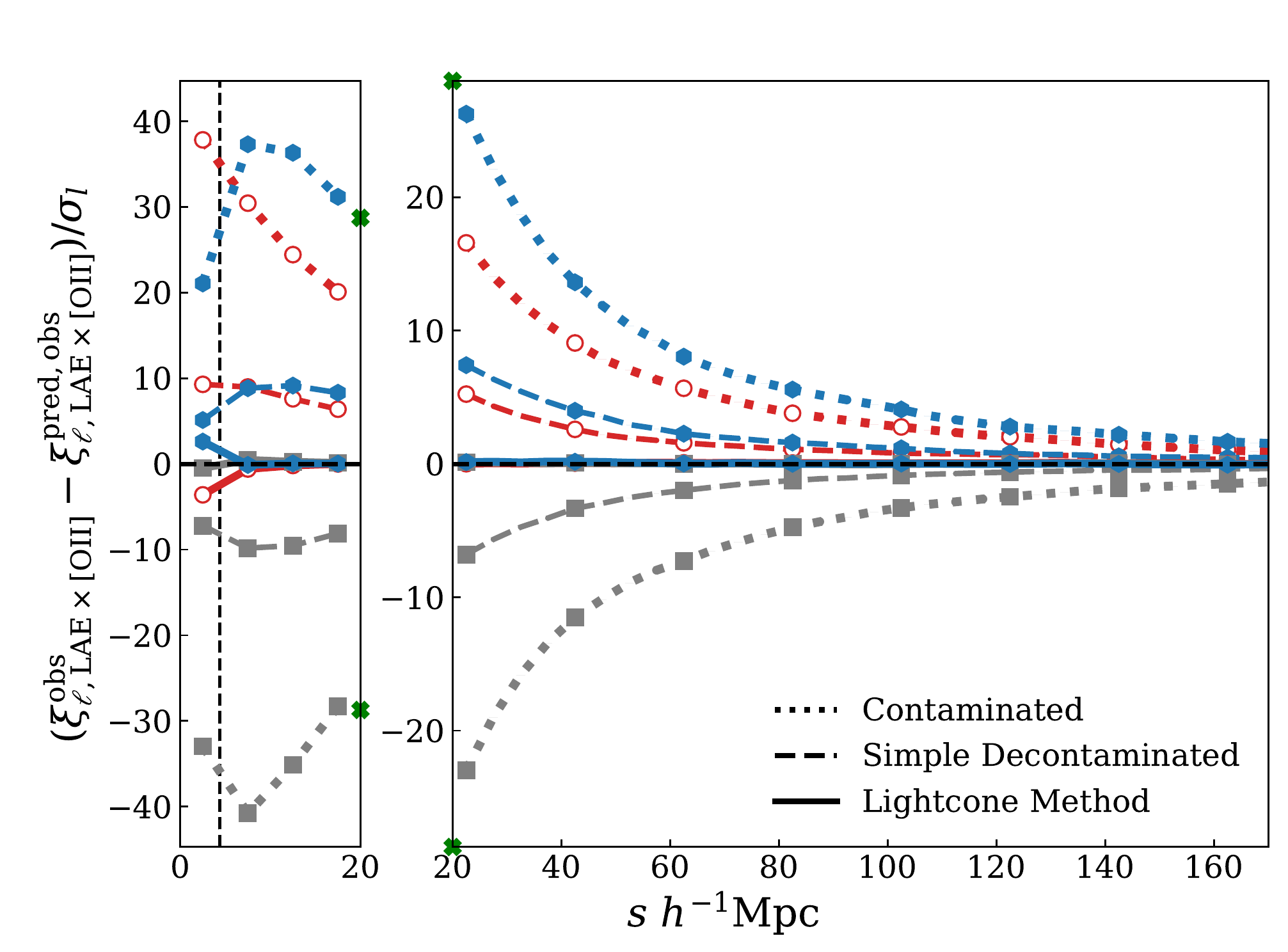} \\
    \end{tabular}
    \caption{The same relations shown in Figure~\ref{fig:full_range_xi}, except now the redshift is restricted to the range over which \oii emitters are simulated ($z_{\oii}>0.05$). The results are very similar, though the residual spurious cross correlation signal seen when ignoring the redshift dependence on the contamination is decreased. }
    \label{fig:xi_zc}
\end{figure*}

\subsection{Lightcone decontamination}
\label{sec:lc_decon}
In this section we apply our new method of decontaminating the samples while accounting for lightcone/redshift effects. 
As with the simple 
method, we will start by assuming perfect knowledge of the purity and contamination, i.e. $f(z)$ and $f_{\oii}(z)$, and use the redshift dependent contamination fraction measured from the random catalogues (i.e., Figure~\ref{fig:purityz}). The results of this are the solid lines in Figure~\ref{fig:full_range_xi}.

In the low contamination LAE sample auto-correlation function we see little meaningful difference when compared to the simple
method. Both return clustering multipoles with very little evidence of systematic bias.   On the other hand, 
in the high contamination sample, the lightcone-based contamination does a better job at correcting the multipoles.
Down to a scale of $20~h^{-1}$Mpc the new method returns measurements with a bias less than $\sim 0.25\sigma$.

The differences between the lightcone model predicted and measured cross-correlation functions show an even greater improvement over the simple method. 
In both the low and high contamination scenarios, the new method accounts for the spurious cross-correlation signal leaving less than $\sim 1\sigma$ residuals at all scales greater than twice the cell size of the LAE simulation box.

The improvements seen in our approach support the idea that the residual, biased signals seen when using simple decontamination come from applying it to clustering measurements without accounting for the significant redshift evolution of the projected clustering and the contamination fraction within the redshift bin.

\subsection{Restricted redshift range}
\label{sec:red_zrange}
The redshift range studied so far, $1.9<z<3.5$, includes a volume in which we expect there to be no \oii emitters.  This is strictly true at redshifts $1.90 < z_{\lae} < 2.06$, since at these redshifts, \oii $\lambda 3727$ would need to be blue-shifted to be confused with Ly~$\alpha$.  As mentioned, for $z_{\lae} < 2.22$, the very small redshift of the \oii emitters ($z_{\oii}<0.05$) would likely allow their classification via their physical sizes and appearance on broadband images.  Thus, 
while studying the full redshift range of HETDEX is a perfectly valid approach, we would also like to see what would happen if we restricted measurements to the range over which \oii galaxies are included in our simulations ($z_{\oii}>0.05$). Removing the redshift range over which the LAE sample is pure means the contamination fraction for LAEs increases to 
$6.8$ per cent for the high contamination sample (defined by $P_{\lae}>0.15$) and $1.8$ per cent for the low contamination case. 

In Figure~\ref{fig:xi_zc} we show results for the redshift range $2.22<z<3.5$. As before the plots with auto-correlations show the difference between the (de)contaminated measurements and those from the corresponding pure catalogues with the same redshift range. In the new case, the contaminated cross correlation function looks the same as for the full redshift range. However, the simple decontamination procedure, which ignores the redshift dependence of the purity within the redshift bin, works better than for the full $z$ range. Nonetheless, it can be seen from Figure~\ref{fig:xi_zc} that even when cutting out redshifts with the most dramatic changes in purity and contamination, not accounting for lightcone effects can still cause biases. 
In the lower contamination case, the  simple decontamination leaves a $\sim 2\sigma$ biased cross-correlation monopole at separations $s>20\,h^{-1}$\,Mpc. This bias increases to $\sim 5\sigma$ for the higher contamination case. The auto-correlation also displays significant biases in the high contamination case if redshift effects are ignored.  The hexadecapole shows a systematic bias even at large scales of up to $\sim 1\sigma$, while the monopole shows a $\sim 2\sigma$ bias at the resolution limit of the simulation. 

In contrast, as for the full redshift range analysis, the new lightcone based decontamination
method returns very close to the true auto-correlation function down to twice the cell size of the LAE box, and also accurately predicts the cross-correlation, with only tiny insignificant residuals, over the same range of scales.
\begin{figure}
    \centering
    \includegraphics[width=0.45\textwidth]{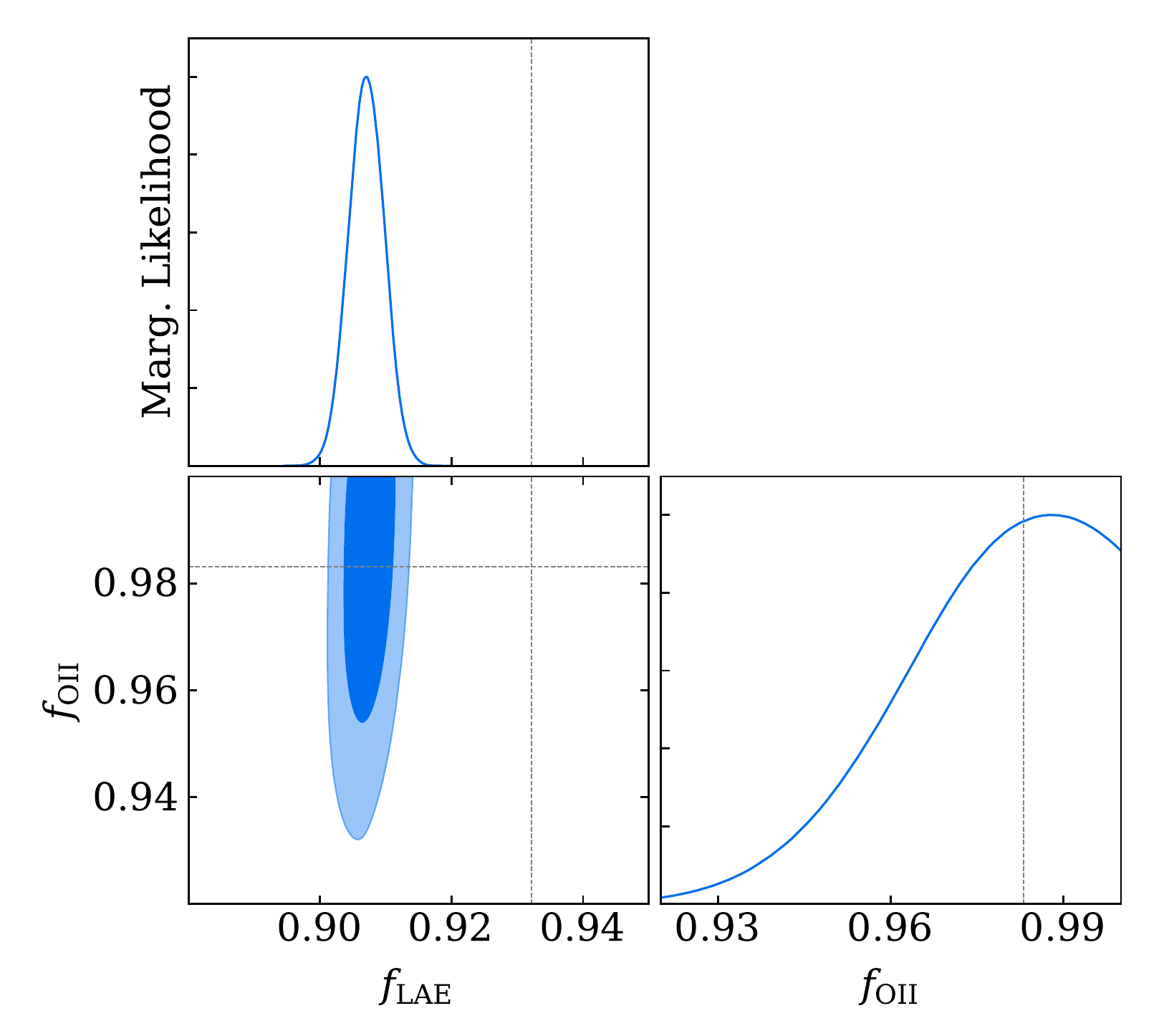}
    \caption{The MCMC 68\% and 95\% contours for the simple decontamination method fits to the contaminated cross-correlation multipoles for the high contamination, $P_{\lae}>0.15$ sample.  The plots assume a single parameter purity model for LAE and \oii galaxies. The dotted lines show the true purity measured directly from the mock catalogues. The normalized and marginalized 1D likelihoods are also shown.}
    \label{fig:chains_matrix}
\end{figure}

\begin{figure*}
    \centering
    \includegraphics[width=0.83\textwidth]{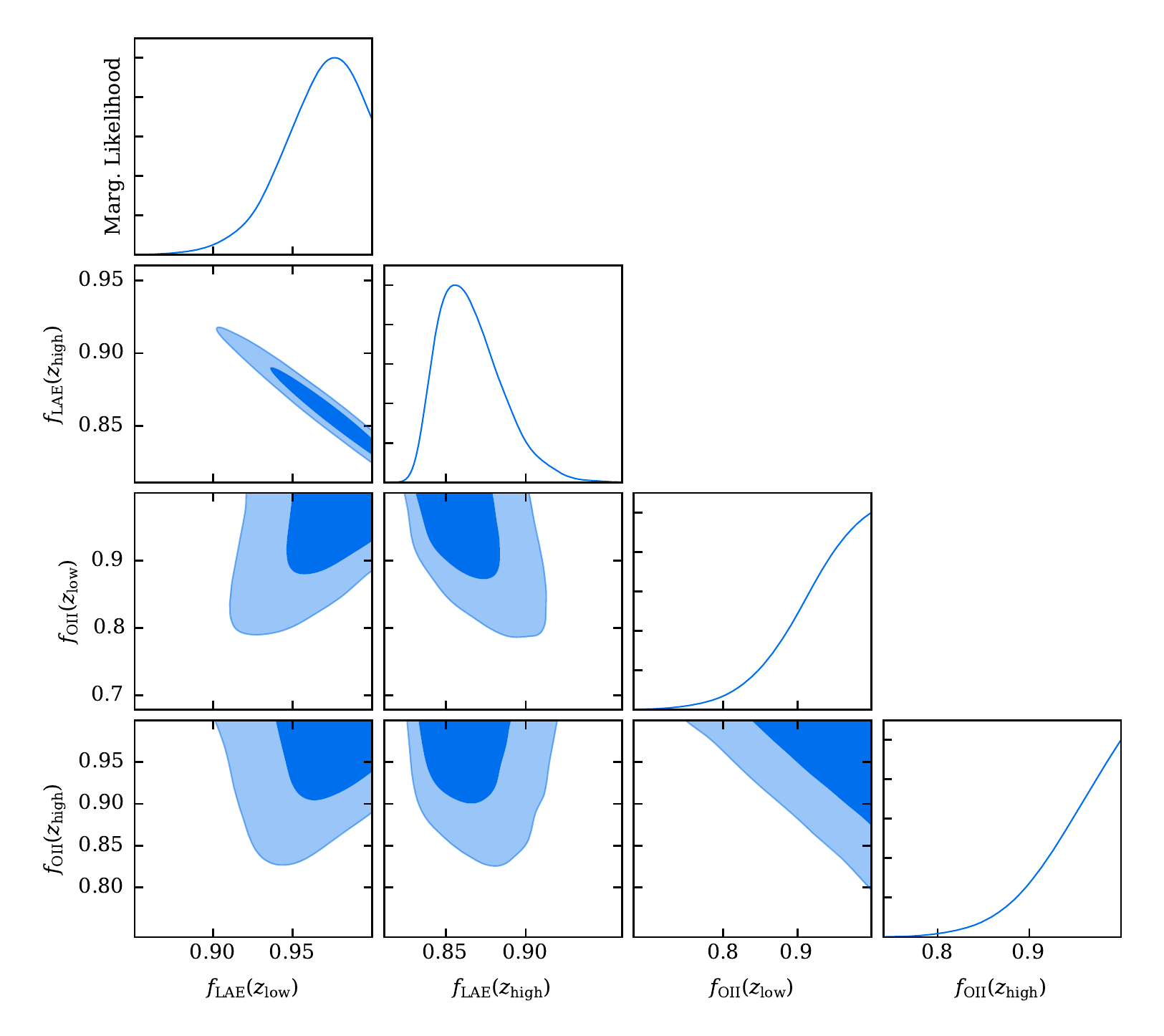}
    \caption{The 68\% and 95\% MCMC contours derived by fitting the cross correlation function of the mock catalogues using a two parameter model for the purity to the LAE and \oii samples. These results are from our lightcone decontamination method. The data are for the high contamination $P_{\lae} >0.15$ sample, which has an average contamination rate of 7\%. The normalized, marginalized 1D likelihoods are also shown.
    A redshift dependence of the contamination of the LAEs is clearly detected.}
    \label{fig:chains015}
\end{figure*}

\begin{figure*}
    \centering
    \begin{tabular}{cc}
      \includegraphics[width=0.45\textwidth]{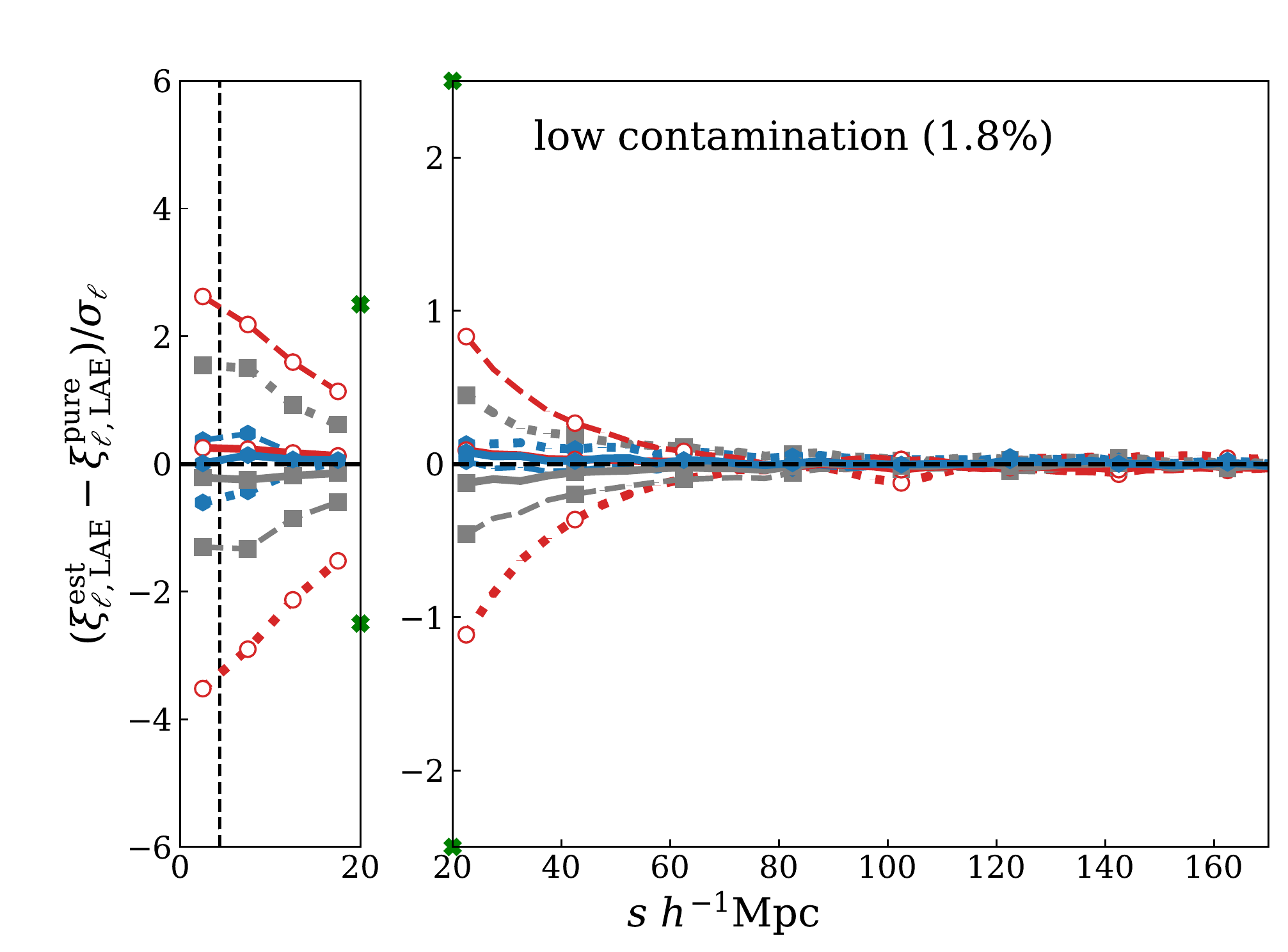} & \includegraphics[width=0.45\textwidth]{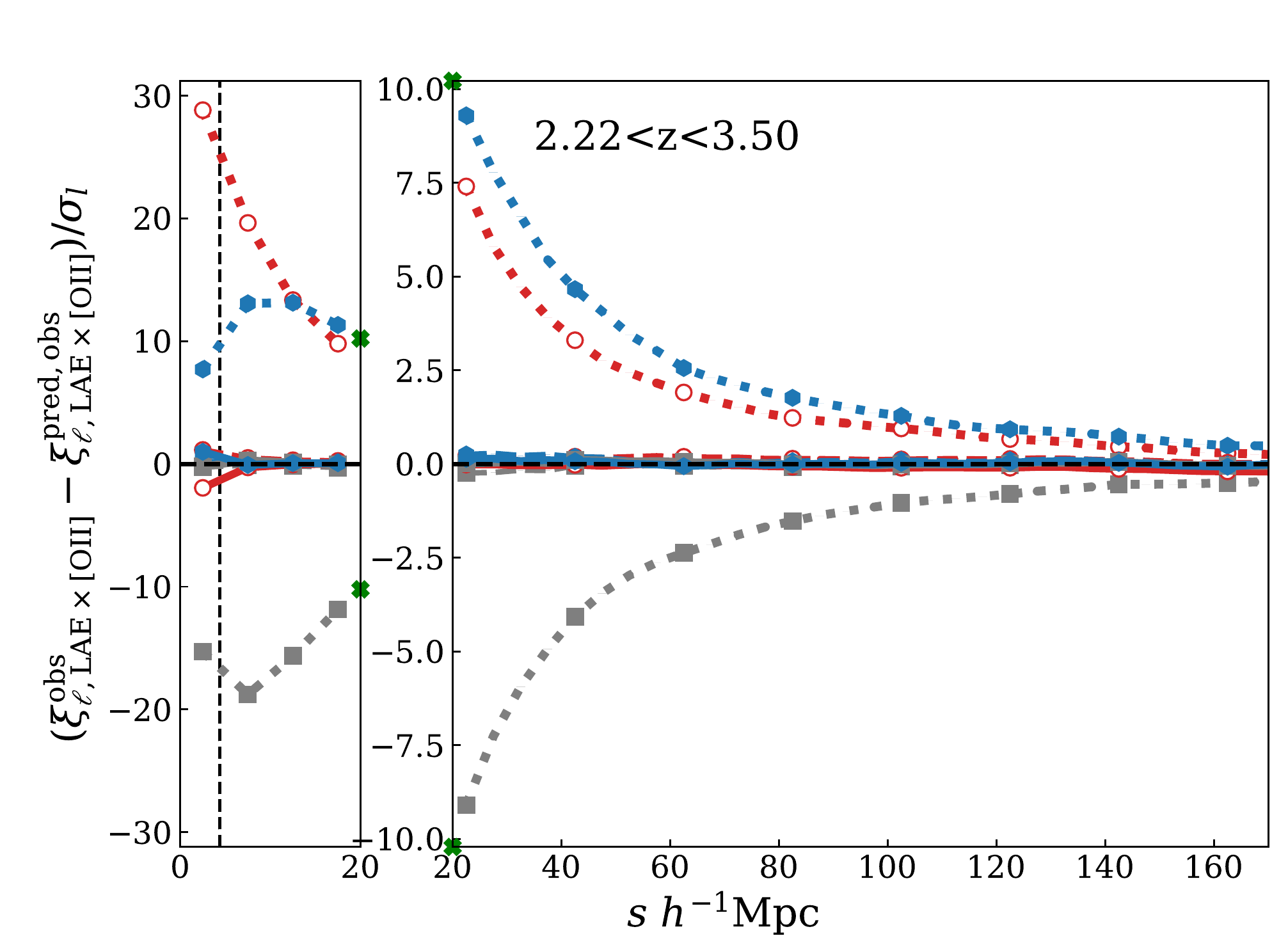} \\
      \includegraphics[width=0.45\textwidth]{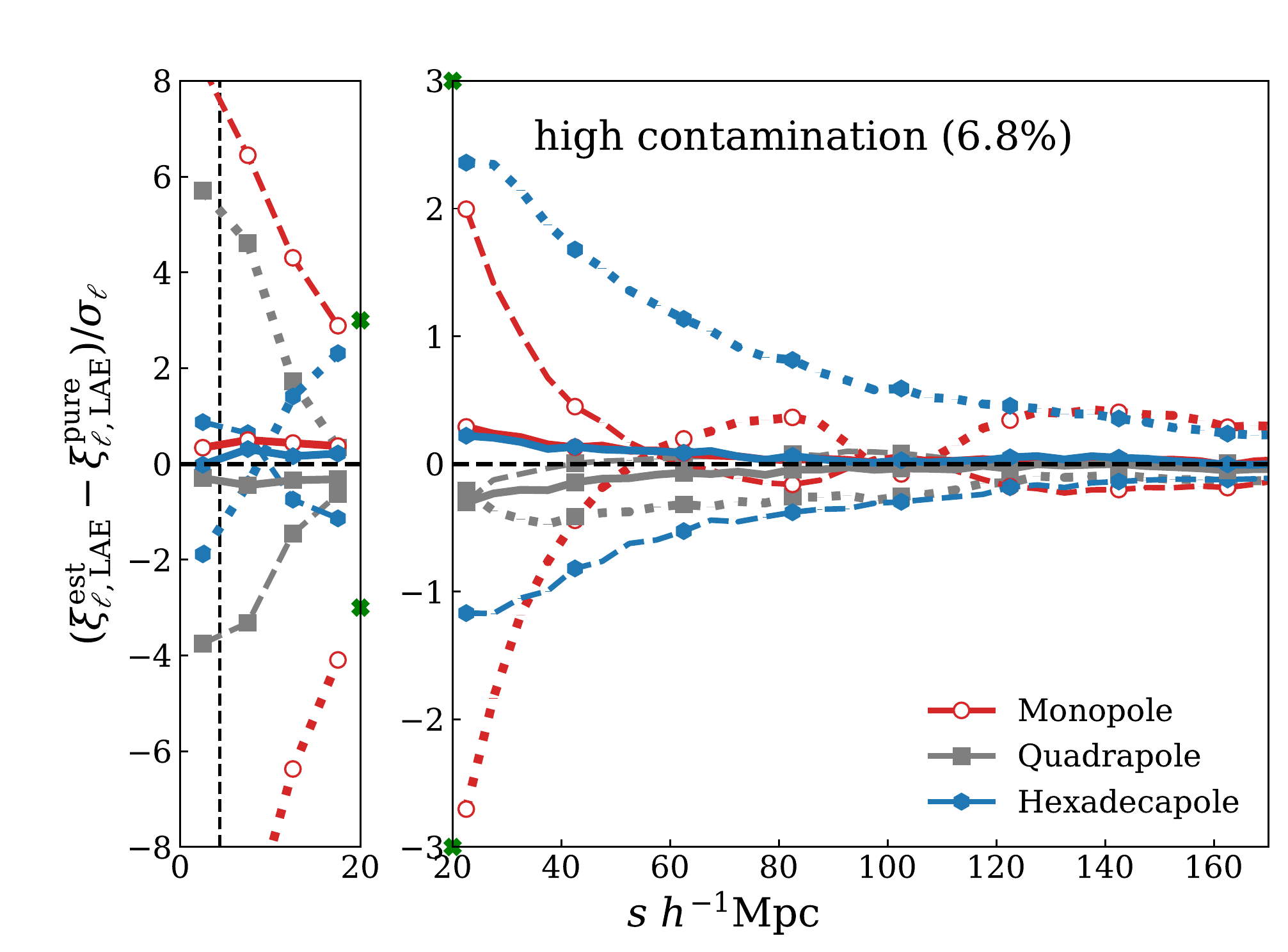} & \includegraphics[width=0.45\textwidth]{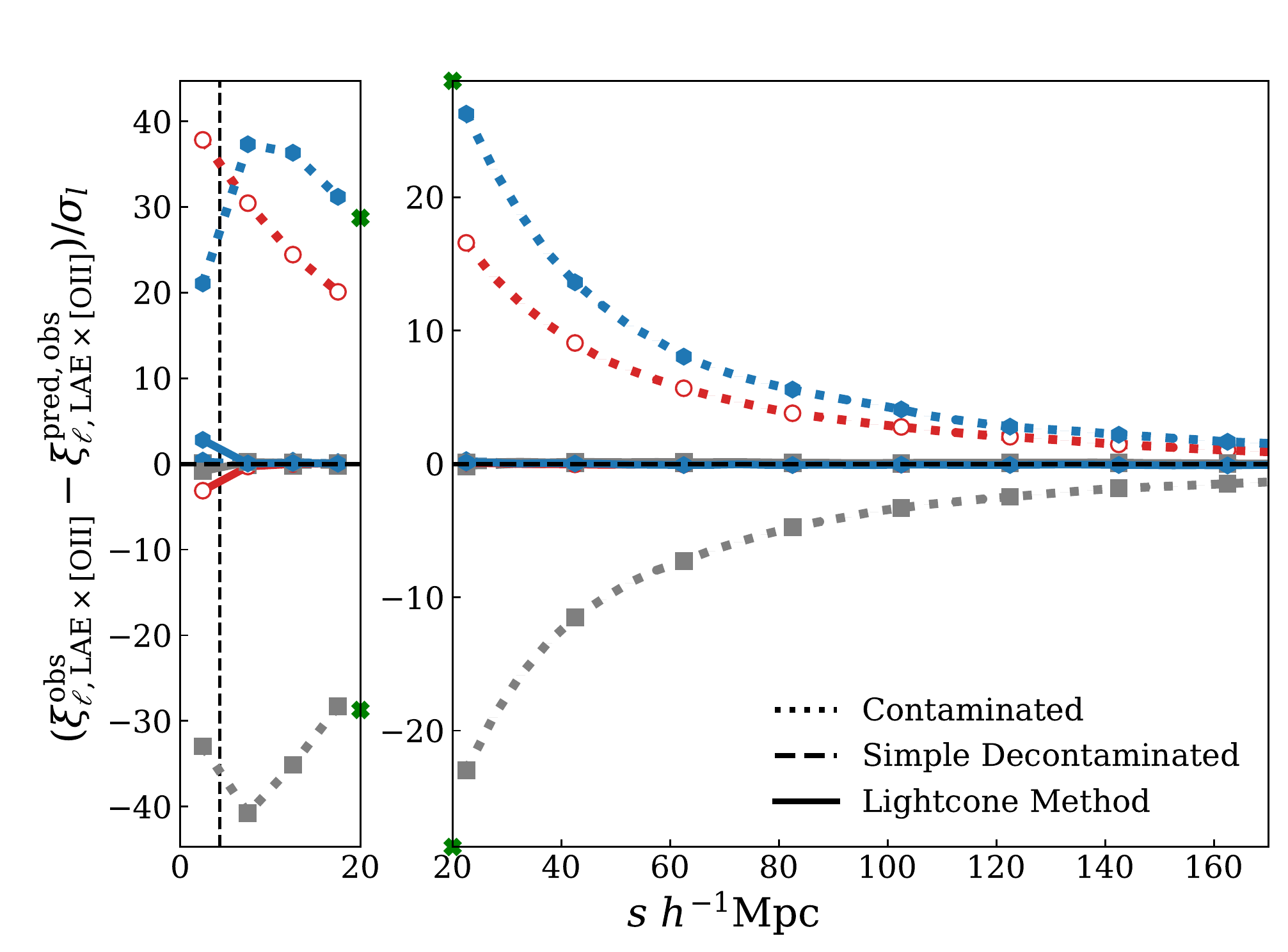} \\
    \end{tabular}
    \caption{This same plot as in  Figure~\ref{fig:xi_zc}, except now instead of using the true purity as a function of redshift, the purity values are determined by fitting the cross correlation functions. In this case all decontamination methods are forced toward returning a zero residual cross-correlation signal in order to minimize the $\chi^2$. However, if we ignore the redshift dependence of the contamination, the best-fitting contamination value is too high; this error propagates directly into additional biases in the decontaminated LAE auto-correlation function. The new biases are most noticeable in the higher contamination results shown on the left panel of the bottom row. The new lightcone approach however works well for both the auto- and cross- correlation functions.}
    \label{fig:decon_bf_015}
\end{figure*}

\section{Fitting the contamination}
\label{sec:fit}
\subsection{Fitting model and technique}
The work so far has assumed we have perfect
knowledge of the contamination. We now attempt to fit for the contamination 
by minimising the residual cross-correlation function, which as mentioned previously, should 
be zero for the case of perfect decontamination as the LAE and \oii samples are in completely separate volumes. We do this using both the simple method and 
our lightcone-based approach to decontamination. We minimise the residual 
difference between the observed cross correlation multipoles and the predicted cross-correlation multipoles evaluated using equations~(\ref{eqn:resid_old}) and (\ref{eqn:deconx}) for the simple and lightcone methods respectively, 
\begin{equation}
	\chi^2 = \left({\bm{\xi}}_{\lae\times\oii}^{\mathrm{obs}} - {\bm{\xi}}_{\lae\times\oii}^{\mathrm{pred, obs}}\right)^T\mathbfss{C}^{-1}\left({\bm{\xi}}_{\lae\times\oii}^{\mathrm{obs}} - \bm{\xi}_{\lae\times\oii}^{\mathrm{pred, obs}}\right),
\end{equation}
where ${\bm{\xi}}_{\lae\times\oii}$ is a vector of all the decontaminated multipole measurements, with superscripts indicating the observed (`obs') and predicted (`pred, obs') cross-correlation functions due to contamination, and $\mathbfss{C}$ is the covariance matrix  of `observed' cross-correlation functions that we measure from the mock realisations. The data to which we fit our model is the mean of the 1000 measured cross-correlation functions, but we use the covariance matrix, $\mathbfss{C}$, appropriate for a single realisation. This means our results will have the reported
uncertainty appropriate for
a single HETDEX realisation, but 
they will be centered much closer to the best model description than statistically likely for real data. 

In the lightcone method for the redshift dependence of the decontamination we use a two parameter model for each sample, where we fit the purity at the high and low edges of the redshift bin $f(z_{\mathrm{low}})$ and $f(z_{\mathrm{high}})$ for both LAEs and \oii galaxies, corresponding to four parameters in total. We linearly interpolate between the two purity fractions to return purity values for the intermediate redshifts. This is motivated by considering the simplest model that could fit the redshift dependent contamination fractions shown in Figure~\ref{fig:purityz}.

We need models of the LAE and \oii clustering to make predictions for the cross-correlation for given contamination parameters. One approach could be to use the measured LAE and \oii correlation functions, decontaminated using the model whose $\chi^2$ is being evaluated. Since our aim is to present a proof of concept, and highlight the sensitivity of these statistics to the redshift dependent $f(z)$, we avoid 
this extra complication by using the true \oii and \lae\ correlation functions, taken as the mean measurement of the 1000 pure mocks of the Fall field. For the OII correlation function in the simple method we use 
the mean correlation function measured from 1000 pure OII catalogues using \lae\ redshifts. 

If one did estimate auto-correlation functions from the data, then their errors would have to be accounted for in the fitting routine. For the simple method one could use equation A16 of \citet{awan20}, which is an expression for the covariance of the decontaminated auto- and cross-correlation functions that propagates the errors on the observed auto- and cross- correlation functions. For the lightcone 
method, we argue we can avoid these issues in future work by including models of the \lae\ and \oii correlation functions in the fitting (see more discussion in section~\ref{sec:conc}). As we only use the covariance of the cross correlation function in our fits, it means the
uncertainties in our parameter estimates correspond to the unrealistically
optimistic scenario of perfect measurements or knowledge of the \lae\ and
\oii auto-correlation functions. 

We fit the sample in the restricted redshift range $2.22<z_{\mathrm{LAE}}<3.50$ since we assume all lower redshift \oii emitters can be correctly classified by imaging. We note even in the restricted redshift range there is an area that in practice has 100 per cent LAE sample purity, where many other \oii emission lines also lie in the HETDEX spectral range.  Because of this, and the presence of other sharp features in the purity versus redshift relation, our simple straight line model for the contamination is not an ideal model of $f(z)$.  Nevertheless,
we shall see it does a reasonable job of describing the behavior of contaminants.

To explore the posterior of the contamination model parameters we use the software package {\sc cobaya} \citep{torrado2021}, which uses the MCMC sampler of \citet{lewis2002} and \citet{lewis2013}. We use a flat prior between $0.6<f<1.0$ for the purity parameters, and compute the likelihood as $\mathcal{L}=\mathrm{exp}(-\chi^2/2)$. We ran eight chains for each fit and set the convergence criteria in {\sc cobaya} to require that a value of the Gelman-Rubin statistic \citep{gelman1992, brooks1998}, modified
as described in \citet{lewis2013}, of $R-1=0.005$ is achieved (the {\sc cobaya} parameter {\sc rminus1\_stop}). The smallest separation bin we use is $15.0<s[\,h^{-1}$\,Mpc$]<20.0$. Even though we assume perfect knowledge of the LAE and \oii auto-correlation functions, we 
avoid using
smaller scales in the fit.  This mimics a scenario where a model of the auto-correlation function might not work sufficiently well on small, more non-linear scales. The MCMC contour plots, confidence intervals and best-fitting parameters were derived using the {\sc getdist} Python package \citep{lewis2019}.

\subsection{Results of the fit from the simple method}
The simple method results of our MCMC fits to the cross-correlation function of the low contamination LAE sample yield the sample purities $\tilde{f}=0.975 \pm 0.003$ for LAEs and $\tilde{f}_{\oii}=0.96 \pm 0.03$ for \oii emitters. The tildes specify quantities estimated from the MCMC fitting. 
We quote the best-fitting value and the range between the maximum and minimum value within 68 per cent of the highest likelihood weighed chain values. In section~\ref{sec:purityc} we explain why we do this rather than quoting the marginalised mean and standard deviation. 
These values can be compared to the total purity and contamination of the whole sample. The LAE purity has a slight bias with respect to the true purity of the sample ($f=0.982$), while the \oii purity value agrees with the true value of $f_{\oii}=0.956$ (these true values are exact to the given number of significant figures.) Note that since we fit to the mean of the mock measurements, the detection of this bias in LAE purity is more significant than its size. 

The biases from the 
simple method of fitting for contamination are small in the low contamination sample. In the high contamination case (where 6.8 per cent of the LAE sample in the restricted redshift range are \oii emitters), the bias on the LAE contamination is larger. We plot the results of the MCMC chains 
in Figure~\ref{fig:chains_matrix}. We can see from the figure that the LAE purity is much better constrained than the \oii purity; the best-fitting 
LAE and \oii purity values are $\tilde{f}=0.907\pm0.004$ and $\tilde{f}_{\oii}=0.99\pm^{0.01}_{0.03}$, respectively.  For comparison, the true purity values are $f=0.932$ and $f_{\oii}=0.983$. In this case the LAE purity measurement is biased much lower than the truth, this is 
because the true purity values leave spurious cross-correlation signals (see Figure~\ref{fig:xi_zc}), so a better $\chi^2$ is given by biased parameters that return a smaller cross-correlation signal.

\subsection{Results of the fit using the lightcone method}
We now turn our attention to the new lightcone decontamination model, which uses two parameters to describe the redshift dependence of the sample purity. The results of our MCMC fitting of the cross-correlation function for the high contamination sample are given in 
Figure~\ref{fig:chains015}. We can see a degeneracy between the low and high redshift purity limits. Lower high-redshift completeness can be somewhat mitigated by higher low-redshift completeness. However, this degeneracy is not complete, and more contamination at higher redshifts is favoured. The \oii emitters, on the other hand, are consistent with 100 per cent purity and there is no significant detection of any redshift dependence of the contamination. 

The MCMC chains of the low contamination sample also show the contamination of the LAEs is detected.  In this case, however, no redshift dependence on purity is found. For the \oii sample, 100 per cent purity is disfavoured but there is a degeneracy in the $f_{\oii}(z_{\mathrm{low}})$ and $f_{\oii}(z_{\mathrm{high}})$ parameters, which means the slope of the purity-redshift relation is not well constrained. Unlike in the 
simple method, there is no true value to 
compare the best-fitting parameters to (or to include in Figure~\ref{fig:chains015}), as a straight line is not a perfect model of the true $f(z)$ or $f_{\oii}(z)$. Instead, in section~\ref{sec:purityc} we compare the straight-line fit from this method directly to the true purity versus redshift relations.

\subsection{Constraints on the purity}
\label{sec:purityc}
We now wish to visualize our constraints in a
plot of purity versus redshift.
The flat priors on the purity at the ends of the redshift range result in non-flat priors on the derived values of purity at intermediate redshifts. For example, at the center of the redshift range a contamination fraction in the middle of the prior range is favoured, as there are more allowed values of $f(z_{\mathrm{low}})$ and $f(z_{\mathrm{high}})$ that produce a purity crossing that point.
In addition, we have the added complication that the expected purity is very close to the priors we use, and those priors cannot be expanded without including nonsensical values of purity (i.e., $>1.0$).  Therefore
to minimise the effects of priors when plotting the purity constraints of
Figure~\ref{fig:purityz}, we do not plot the weighted mean and standard deviation of the chains. Instead we show the best-fitting parameters as a dotted line, and
the maximum and minimum values of purity in 68 per cent of weighted chain values with the highest likelihoods as a shaded region. We compute these parameters using the likelihood functions in {\sc getdist} \citep{lewis2019}.

We can see in Figure~\ref{fig:purityz} that for both of the P$_{\lae}$ cuts, and for both the LAE and \oii galaxy samples, the best fitting parameters qualitatively reproduce the slope and amplitude that would be expected for a straight line fit to the more complex behaviour of the true purity. The true purity as a function of redshift is nearly always within the minimum and maximum range defined by the 68 per cent of chain positions with the highest likelihood. We do however see the true purity is slightly outside the region at the two redshift extremes. This is reasonable, as at those positions the true purity deviates most from the straight line. It is possible the fits could be improved by adding additional parameters to the model to mimic the sharp drops in purity. However, we will see in section~\ref{sec:using_fit_decon} that this detailed modelling is unnecessary to decontaminate the clustering measurements for the scenarios considered here. It is also unclear whether the data are really constraining enough to warrant additional model parameters. 

In general, we see that the LAE purity is better constrained than the \oii purity. A possible cause of this is the choice of measuring the cross correlation function using LAE redshifts for both samples. This mapping between observed wavelength and redshift means that the LAE contamination in the \oii sample has a correlation function that is fixed with redshift.  This allows
us to remove it from the integral in equation~(\ref{eqn:deconx}). Since the LAE clustering term is independent of redshift, the sensitivity to the redshift dependence of purity in equation~(\ref{eqn:deconx}) comes from the $f_{\oii}(z)(1 - f(z))$ term. This term is most sensitive to changes in $f(z)$, since $f_{\oii}(z)$ is always close to one but $(1 - f(z))$ is always very small. 
One way to better constrain the \oii purity versus redshift could be to measure the cross-correlation using \oii redshifts for all sources. Then the projected LAE clustering would evolve over the redshift range of \oii, making the cross correlation more sensitive to $f_{\oii}(z)$. 
Note this approach is not guaranteed to give a strong constraint on the \oii purity, as the \lae\ sample has an intrinsically weaker correlation function, which may get even weaker when projected to lower redshifts. Since we are mainly interested in the contamination of the LAE correlation function, and $f(z)$ is the only purity term appearing in equation~(\ref{eqn:autodecon}), we leave further study of this to future work.

\subsection{Using the fitted contamination}
\label{sec:using_fit_decon}
In section~\ref{sec:results} we assume we have perfect knowledge of the purity as a function of redshift. Here we use the purity we have fit from the cross-correlation function, to see if it can be used to give unbiased,
decontaminated measurements of the auto correlation functions. We decontaminate the measurements from the two fields separately using the estimated parameters from the combined field, and then combine them after the decontamination \citep[following, e.g.,][]{white2011}. To test if combining before or after decontamination makes a difference in our mocks, we tried combining the two fields before decontaminating for one of the scenarios, specifically the high contamination sample using the simple decontamination method. We found
results that agree closely, confirming the order of combining and decontaminating is not important for our mocks. In the future, further tests could be carried out to find if this is also the case with the real HETDEX data. 

In Figure~\ref{fig:decon_bf_015} we show the results of using the best-fitting contamination parameters. We can see that these parameters return much smaller residual cross-correlation signals than the true parameters for the simple approach that ignores redshift effects. This is because it is able to fit the extra cross-correlation signal with artificially low purity values. However, the LAE auto-correlation function shows the danger in this, as the incorrect inferred purity values results in even more biased results than when using the true values. The monopole from the high-contamination sample can have a $\sim 2\sigma$ bias even at larger scales ($s>20\,h^{-1}$\,Mpc) and the bias gets even worse at small scales. The low-contamination sample monopole shows a $\sim 1\sigma$ bias at large scales, increasing to $\sim 2\sigma$ at small scales. 

In contrast, our new lightcone method of accounting for the redshift dependence of contamination works well for both the auto and cross-correlation functions, yielding auto-correlation measurements with only a $\sim 0.3\sigma$ bias at large scales ($s>20\,h^{-1}$\,Mpc) and biases smaller than the statistical error down to twice the cell size of the simulation box. The results 
from the fitted contamination are slightly worse than the results from using the true purity versus redshift, but this is to be expected given the limitations of the model parameterisation and the added uncertainties from the fit. These results show this method could potentially be used to 
both fit for contamination and correct clustering measurements.

It is important to note that in Figure~\ref{fig:decon_bf_015} we have the benefit of a best-fitting purity measured from
our large number of mock catalogues. In the real data, there 
will be some statistical error associated with the best-fit 
purity parameters that would have to be propagated into the final results.

It should also be noted that an alternative method of deriving the $f(z)$ to the one presented here is to use
the simple decontamination method in narrow
redshift bins where the contamination and projection parameters are roughly constant. This could have benefits, such as avoiding potentially
 losing information by not integrating over redshift, and not needing a model of the \lae\ or \oii correlation function. On the other hand, using narrow redshift bins would make the individual measurements noisier. We leave
a comparative study of these approaches for
later work. We also highlight regardless of how the
$f(z)$ is derived, the lightcone formalism allows one to optimize the size of redshift bins for the cosmology measurements without having to be restricted by the requirement of avoiding projected interloper clustering evolution.

\FloatBarrier
\section{Conclusions}\label{sec:conc}
We generated 1000 mock catalogues of the HETDEX survey, which include clustering, redshift-space distortions, redshift-dependent noise and a realistic selection function. We used a reformulated version of the probabilistic classifier of \citet{leung} to generate catalogues of LAEs with  realistic, redshift-dependent \oii galaxy contamination, and considered two scenarios which bracket the expectations for HETDEX: low contamination (1-2 per cent) and high contamination  (5-7 per cent). These catalogues were used to explore  the impact of the redshift dependence of the contamination fraction and the correlation function of the contaminants on the observed correlation functions.

The mock catalogues show that existing methods of decontamination such as \citet[][]{awan20}, and other methods which do not account for redshift evolution of the interlopers within a redshift bin, should not be directly applied to clustering measurements from a survey such as HETDEX, unless the analysis is restricted to using redshift bins that are narrow enough for the evolution effects to not be important - which is unlikely to be ideal for the HETDEX cosmological analysis. This is because in HETDEX both the \oii galaxy contamination fraction and the projected \oii clustering vary with redshift. Although in the low contamination cases we 
consider, such methods may be effective enough, when the contamination is larger (with misclassified fractions of 5-7 per cent) biases appear. In the auto-correlation function these biases can be larger than the statistical error of HETDEX\null. Moreover, the biases in the cross correlation signal are even stronger, with a spurious signal of more than $5\sigma$ of the expected statistical noise being left after subtracting the expected cross correlation from contamination. A biased, decontaminated cross-correlation function is not a problem 
for HETDEX in itself. However, we have also shown the inability of the simple 
decontamination method to correctly account for the spurious cross-correlation signal results in biases in the inferred contamination fractions from fits to the observed, raw cross-correlation functions. If these incorrect contamination fractions are used to decontaminate the auto-correlation function, additional biases will be propagated into the measurements used to fit cosmological parameters. 

We present a method to account for the redshift effects that can be applied when there is no true cross correlation between the pure samples of the target galaxies and the contaminants. Our method combines the literature approach to decontamination with the models of correlation functions integrated along a lightcone given in \citet{yamamoto1999} and \citet{suto2000}. This methodology is needed in scenarios like HETDEX, where the correlation function of the \oii interlopers evolves rapidly due to projection effects.
Accounting for the lightcone effects gives a much better model of the cross correlation, and it also produces decontaminated auto-correlation functions that agree with the pure measurements to an accuracy much smaller than the statistical noise down to $20h^{-1}$\,Mpc. Although we
formulate this work for the correlation function, our findings should also apply to the power spectrum.

The work on this topic is not complete. The method we have developed is an improvement over existing methods, but we still have to assume that the 
true clustering of LAEs and \oii galaxies does not evolve with redshift. Allowing for evolving LAE and \oii correlation functions is possible in the framework we present however, and such an evolution could be constrained with auto-correlation function measurements from the data. We
also mention once more that we always compute the projected clustering using distortion parameters assuming a true cosmology. As advocated by \citet{addison2019}, future work could consider the impact of this limitation.

Although our decontamination method itself relies only on interpolating over the observed
correlation functions and the assumption of a fiducial cosmology for computing the distortion parameters,
our method of fitting the contamination assumes perfect knowledge of the LAE and \oii  auto-correlation functions. One way to make our experiments with fitting the contamination applicable to real data would be to develop
an approach that simultaneously fits models of the cosmology, bias, and the contamination parameters to both the cross- and auto-correlation functions. The distortion parameters could also be modified to be consistent with the different cosmologies during the fit, solving a further issue. 

Simultaneously fitting contamination, galaxy bias, and cosmology is advocated in e.g., \citet{addison2019} and \citet{grasshorn19}. While their 
approaches assumed single contamination values for
the whole sample, we 
would suggest including our new redshift dependent modelling of contamination.
This is especially true if the contamination fraction of HETDEX is closer to 5 per cent than 1 per cent, and if there are significant differences in the sample purity at different redshifts. Before applying the method to real data, a modelling pipeline which includes redshift-dependent contamination should be tested on more realistic
simulations than our log-normal mocks. These simulations should include possible evolution in the true galaxy correlation function and a better modelling of the clustering and redshift space distortions on non-linear scales.

To summarize, this paper highlights the importance of the redshift dependence of the contamination, presents a method to model these effects, and shows that
such effects should be considered
when decontaminating  clustering measurements from surveys with redshift dependent contamination within the adopted redshift bins. These effects are particularly important when using the cross-correlation function to constrain contamination. 

\section*{Acknowledgements}
The authors acknowledge the feedback from the internal HETDEX referees and the anonymous journal referee.
We acknowledge useful discussions with Humna Awan, Jiamin Hou, Martha Lippich, Andrea Pezzotta, Agne Semenaite, Mart\'in Crocce,  Rom\'an Scoccimarro and the HETDEX cosmology science working group. Henry S. Grasshorn Gebhardt is a NASA Postdoctoral Program
Fellow. KG acknowledges support from NSF-2008793.
EG was supported by the Department of Energy via grant DE-SC0010008. EK's work was supported in part by the Deutsche Forschungsgemeinschaft (DFG, German Research Foundation) under Germany's Excellence Strategy - EXC-2094 - 390783311. DJ was supported at Pennsylvania State University by the NASA ATP program (80NSSC18K1103). 
We acknowledge the use of the Python libraries {\sc matplotlib} \citep{hunter2007}, {\sc astropy} \citep{astropy2013, astropy2018}, {\sc numpy} \citep{harris2020} and {\sc scipy} \citep{virtanen2020}. This research also used {\sc topcat} \citep{taylor2005}, {\sc stilts} \citep{taylor2006} and the GNU Scientific Library (GSL). URL: https://www.gnu.org/software/gsl/

HETDEX is led by the University of Texas at Austin McDonald Observatory and Department of Astronomy with participation from the Ludwig-Maximilians-Universität München, Max-Planck-Institut für Extraterrestrische Physik (MPE), Leibniz-Institut für Astrophysik Potsdam (AIP), Texas A\&M University, Pennsylvania State University, Institut für Astrophysik Göttingen, The University of Oxford, Max-Planck-Institut für Astrophysik (MPA), The University of Tokyo and Missouri University of Science and Technology. In addition to Institutional support, HETDEX is funded by the National Science Foundation (grant AST-0926815), the State of Texas, the US Air Force (AFRL FA9451-04-2- 0355), and generous support from private individuals and foundations.

The observations were obtained with the Hobby-Eberly Telescope (HET), which is a joint project of the University of Texas at Austin, the Pennsylvania State University, Ludwig-Maximilians-Universität München, and Georg-August-Universität Göttingen. The HET is named in honor of its principal benefactors, William P. Hobby and Robert E. Eberly.

The authors acknowledge the Texas Advanced Computing Center (TACC) at The University of Texas at Austin for providing high performance computing, visualization, and storage resources that have contributed to the research results reported within this paper. URL: http://www.tacc.utexas.edu 

This research made use of NASA's Astrophysics Data System Bibliographic Services.

\section*{Data Availability}
 The HETDEX data is currently proprietary, but public releases are planned for the future. The authors will respond to reasonable requests for access to the simulation data used in this paper, so long as no unreleased proprietary data is involved. 



\bibliographystyle{mnras}
\bibliography{bibliography} 




\appendix

\section{Tests of the projection}
\label{sec:appendix}
In this section we further test our modelling of the light cone projection and our use of the distortion parameters $c_{\perp}$ and $c_{\parallel}$. To do this we measure the projected clustering of pure samples of \oii emitters in 1000 of our Fall field mock
catalogues. This means we measure the multipoles of a pure \oii catalogue assuming Ly~$\alpha$ redshifts for everything. To assess how well equation~(\ref{eqn:lc}) 
models the redshift dependent distortion of the \oii density field we compute a prediction of the projected correlation function for the redshift interval $2.22<z<2.5$ via
\begin{equation}
\xi^{{\rm proj}}(s, \mu) = \mathcal{F}[n_{\oii}^{{\rm pure}}(z), \xi^{{\rm proj}}_{\oii}(s, \mu, z)],
\end{equation}
where $n_{\oii}^{{\rm pure}}(z)$ is the number density as a function of LAE redshift of the projected, pure \oii sample. We then compare the multipoles of the result to the multipoles measured from the mock catalogue $\xi^{{\rm mock}}_{\ell}(s)$. 

\begin{figure}
    \centering
    \includegraphics[width=0.47\textwidth]{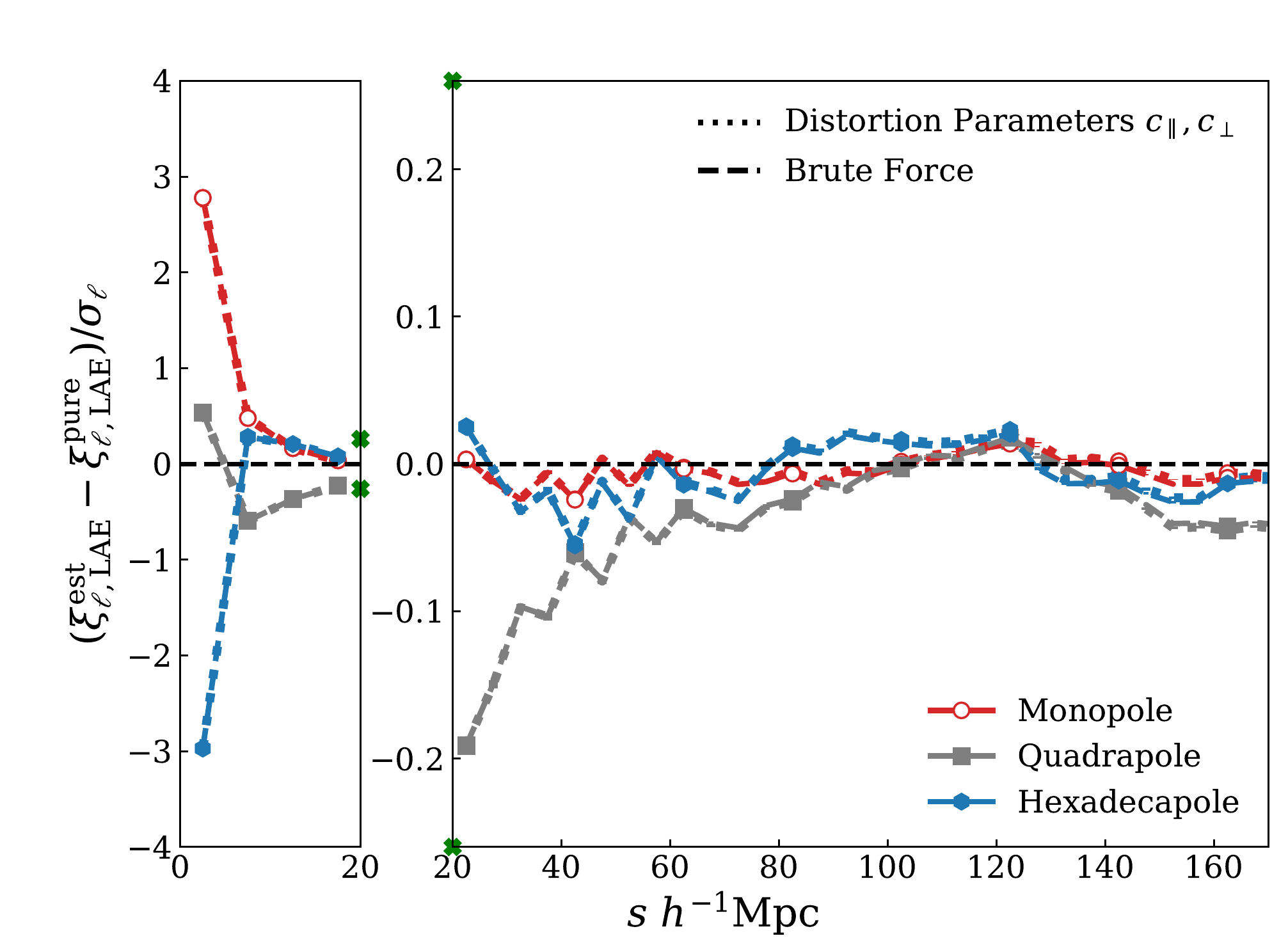}
    \caption{The difference between the mean multipoles measured from
             1000 Fall field mock catalogues of \oii  emitters, analysed using redshifts
             assuming the \oii emitters are actually LAEs, and our prediction of the projected \oii galaxy clustering. Dashed lines use our standard approach with distortion parameters and an integral along the redshift range. The dotted lines 
             replace the distortion parameters with the brute force approach detailed in the text. The distortion parameters and the brute force approach give 
             nearly indistinguishable
             results. The plot is split into two panels to allow a larger dynamic range, in the right panel only every forth data point is marked with a symbol for visual clarity.}
    \label{fig:proj_test}
\end{figure}

The difference between the mock and predicted multipoles of the projected \oii galaxy correlation function is given as dashed lines in Figure~\ref{fig:proj_test}; the measurements are divided by the statistical error expected for one HETDEX Fall-field realisation. We can see only very small ($\sim 0.2\sigma$) differences between the statistical errors down to $s=20\,h^{-1}$\,Mpc, which gives further confirmation of our use of the 
\citet{yamamoto1999} and \citet{suto2000} approach to account for the redshift dependence of the projection parameters. On scales smaller than this, the differences increase, becoming approximately the same size as the statistical error around $s=10\,h^{-1}$\,Mpc. It is unclear why the modeling of the projection does not work perfectly down to the smallest scales, but for the purposes of decontaminating the LAE signal, the
projected \oii clustering is down weighted by the square of the LAE sample contamination.  Thus this small bias
should not affect our results on LAE clustering decontamination. In  section~\ref{sec:fit} where we fit
the cross correlation, we restrict ourselves to larger separations 
which will also mitigate any possible effect. 

In this paper we use a model of  $\xi^{{\rm proj}}_{\oii}(s, \mu, z)$ that relies on distortion parameters, i.e., equation~(\ref{eqn:disto}). These distortion parameters are, however, an approximate model of the effects of the true projection as they do not account for the different
redshifts of the two galaxies in the pair. We therefore also compute a mapping between the $s$ and $\mu$ coordinates using a
brute force method. In 800 uniform bins of LAE redshift between $2.06<z<3.5$ (i.e., the range where a Ly~$\alpha$ emitter has a wavelength greater than 3727~\AA) we generate pairs of galaxies in cartesian coordinates with given values for the true separation $s$, $\mu$ and compute an observed right ascension and declination to a virtual observer. The pairs we generate are always in a plane and use a fixed line-of-sight. We then recompute the cartesian coordinates from our `observed' coordinates but assume \oii  redshifts, and recompute the line-of-sight to
the galaxy pair. By measuring the $r'$, $\mu'$ of this projected pair we generate a look-up table between true and projected coordinates in 400 bins of $r$ and $\mu$. We interpolate over this 3D look-up table to provide an alternative coordinate mapping for our model of the projected \oii clustering. 

The dotted lines in Figure~\ref{fig:proj_test} show the difference of the mock multipoles and the projected multipoles using the brute force look-up table for 
the mapping between LAE and \oii coordinates. There is extremely close agreement between
the predictions using the distortion parameters and the brute force look-up table. This confirms that the distortion parameters $c_{\parallel}, c_{\perp}$, which are advocated by 
several other authors \citep[e.g.,][]{visbal2010, gong2014, lidz2016, pullen2016, leung, grasshorn19}, are an excellent model 
of the true distortion for the HETDEX LAE/\oii confusion scenario.


\bsp	
\label{lastpage}
\end{document}